\documentclass[range]{ar2e}

\usepackage{natbib} 
\bibpunct{(}{)}{;}{a}{}{,}  

\usepackage{lscape} 

\usepackage{ifthen} 

\def\gray{$\gamma$-ray\ }
\def\grays{$\gamma$-rays\ }

\begin{document}

\input epsf.tex    
\input epsf.def   

\input psfig.sty

\jname{}
\jyear{}
\jvol{}
\ARinfo{}

\title{Cosmic-ray propagation and interactions in the Galaxy}

\markboth{Cosmic-ray propagation}{Strong, Moskalenko, and Ptuskin}

\author{Andrew W. Strong$^1$, Igor V. Moskalenko$^2$, Vladimir S. Ptuskin$^3$
\affiliation{
1. Max-Planck-Institut f\"ur extraterrestrische Physik,  Postfach 1312, 85741 Garching, Germany. $aws@mpe.mpg.de$\\
 2. Hansen Experimental Physics Laboratory (HEPL)  and KIPAC, Stanford University, Stanford, CA 94305, U.S.A. $imos@stanford.edu$\\
 3. Institute for Terrestrial Magnetism, Ionosphere and Radiowave Propagation of
the Russian Academy of Sciences (IZMIRAN), Troitsk, Moscow region 142190,
Russia. $vptuskin@izmiran.ru$
}}

\begin{keywords}
energetic particles, gamma rays, interstellar medium,  magnetic fields, plasmas {\it This draft  was made on \today.} 
\end{keywords}

\begin{abstract}

We survey the theory and experimental tests for the propagation of
cosmic rays in the Galaxy up to energies of $10^{15}$ eV.  A guide to
the previous reviews and essential literature is given, followed by an
exposition of basic principles. The basic ideas of cosmic-ray
propagation are described, and the physical origin of  its  processes
are explained. The various techniques for computing the observational
consequences of the theory are described and contrasted. These include
analytical and  numerical techniques.  We present the comparison of
models with data including direct and indirect -- especially gamma-ray
-- observations, and indicate what we can learn about cosmic-ray
propagation.  Some particular important topics including electrons and
antiparticles   are chosen for  discussion.

\end{abstract}
\maketitle

\section{Introduction}

Cosmic rays  (hereafter CR) are almost unique in astrophysics in
that they can be directly sampled, not just observed via
electromagnetic radiation.  Other examples are  meteorites and
stardust.  CR  provide us with a detailed elemental and isotopic
sample of the current (few million years old) interstellar medium not
available in any other way.  It is this which makes the subject
especially rich and complementary to other disciplines.

CR  have featured frequently in Annual Reviews: about 15 articles from 1952 to 1989! 
Reviews include heavy nuclei
\cite{1970ARNPS..20..323S},  
collective transport effects
\citep{1974ARA&A..12...71W}, 
composition 
\citep{1983ARNPS..33..323S} 
and propagation
\citep{1980ARA&A..18..289C}. 
Cox's recent Annual Review `The Three Phase Interstellar Medium' 
\citep{2005ARA&A..43..337C} 
contains much discussion of CR as one essential component of the 
interstellar medium (ISM), but inevitably no mention of their propagation.
Two recent  AnnRev articles 
\citep{2004ARA&A..42..211E,2004ARA&A..42..275S} 
give extensive discussion of the relation of CR to 
turbulence which means we do not try to cover this also.


A basic reference is the book  `Astrophysics of Cosmic Rays'
\citep{1990acr..book.....B} 
which expounds all the essential concepts, 
and is an update of the  classic  `The Origin of Cosmic Rays' 
\citep{1964ocr..book.....G} 
which laid the modern foundations of the subject,
with an updated presentation in
\citep{1976RvMP...48..161G}. 
Good books for basic expositions are
\citep{1972cora.book.....H,1990cup..book.....G} 
and for high energies 
\cite{2004hecr.book.....S}. 
A  basic text emphasizing theory is
\citep{2002cra..book.....S}, 
while the  book `Astrophysics of Galactic Cosmic Rays'
\citep{2002agcr.conf.....D} 
gives an valuable overview of the experimental data and theoretical ideas as of 2001.
The bi-annual  International Cosmic Ray Conference 
proceedings\footnote{the recent conferences are available  via  
the NASA Astrophysics Data System ADS} are also an essential source 
of information, especially for the latest news on the subject. 

Recently a plethora  of  reviews have appeared on the subject of CR above 10$^{15}$ eV,  e.g.
\cite{2000RvMP...72..689N,2005NuPhS.138..465C,2005astro.ph..8014H}; 
on interactions
\cite{2006NuPhS.151..437E}, 
on experiments and astrophysics
\cite{2004AIPC..698..357S}, 
more on astrophysics, propagation and composition
\cite{2004NuPhS.136..147B,2006astro.ph.12247B,2006astro.ph..8085H}, 
and a review of models
\cite{2004APh....21..241H}; 
we  recommend
\cite{2006astro.ph..7109H} 
and the very up to date
\cite{2006astro.ph..8553G,2006astro.ph.11884K}. 
 Therefore this
topic has been excluded here. 
At the lowest energy end, we note that  MeV particles are non-thermal (even if not relativistic) and
must be mentioned  in a review of CR, especially since they are important sources of heating and ionization of the interstellar medium
\cite{1998ApJ...506..329W}. 
As one example of their far-ranging influence, star formation in molecular clouds may be suppressed by CR produced in SNR nearby
\cite{2006ApJ...653L..49F}. 


It is worth distinguishing between two ways of approaching CR propagation:
 either  from the {\it particle} point-of-view, including the  spectrum
and interactions, 
or treating the CR as a weightless collisionless relativistic {\it gas} with pressure and energy and considering it alongside other
components of the interstellar medium
 \cite{2003A&A...412..331H,2006MNRAS.tmp.1221S}. 
Both ways of looking at the problem  are valid up to a point,
 but for consistency a unified
approach  would be desirable and to our knowledge has never been attempted. 
The nearest approach to this is 
\citep{1996A&A...311..113Z,1997A&A...321..434P}. 
Most papers address exclusively one or the other aspect.
The first approach is required for comparison with observations of CR (direct and indirect)
while the second is required for the ISM: stability,  heating etc. 
\citep{2005ARA&A..43..337C}. 


The major recent advances in the field are the high quality measurements of isotopic composition and  element spectra,
and  observations by gamma-ray telescopes, both satellite and ground-based.
Space does not allow any discussion of the observational data here, but the figures give an illustrative overview of what is now available
from both direct (Figs.~1--13) and indirect ($\gamma$-ray) measurements (Figs.~14, 16).
Concerning the {\it origin} of CR,
we follow Cesarsky's 1980 Annual Review
\citep{1980ARA&A..18..289C}  
`we will, for the most part, sidestep this problem.'
Hence we omit CR sources including composition and acceleration;
for supernova remnants as CR sources the literature can be traced back from the most recent H.E.S.S. TeV \gray results
\cite{2006A&A...449..223A}. 
 We also omit solar modulation, galaxy clusters and extragalactic CR.
 We mostly restrict attention to our own Galaxy, but mention important information coming from external galaxies  (via synchrotron radiation).

 We first introduce the theoretical background,
and then consider the confrontation of theory with observation.
A number of particular topics are selected for further discussion.

\begin{figure}
\centerline{
\psfig{figure=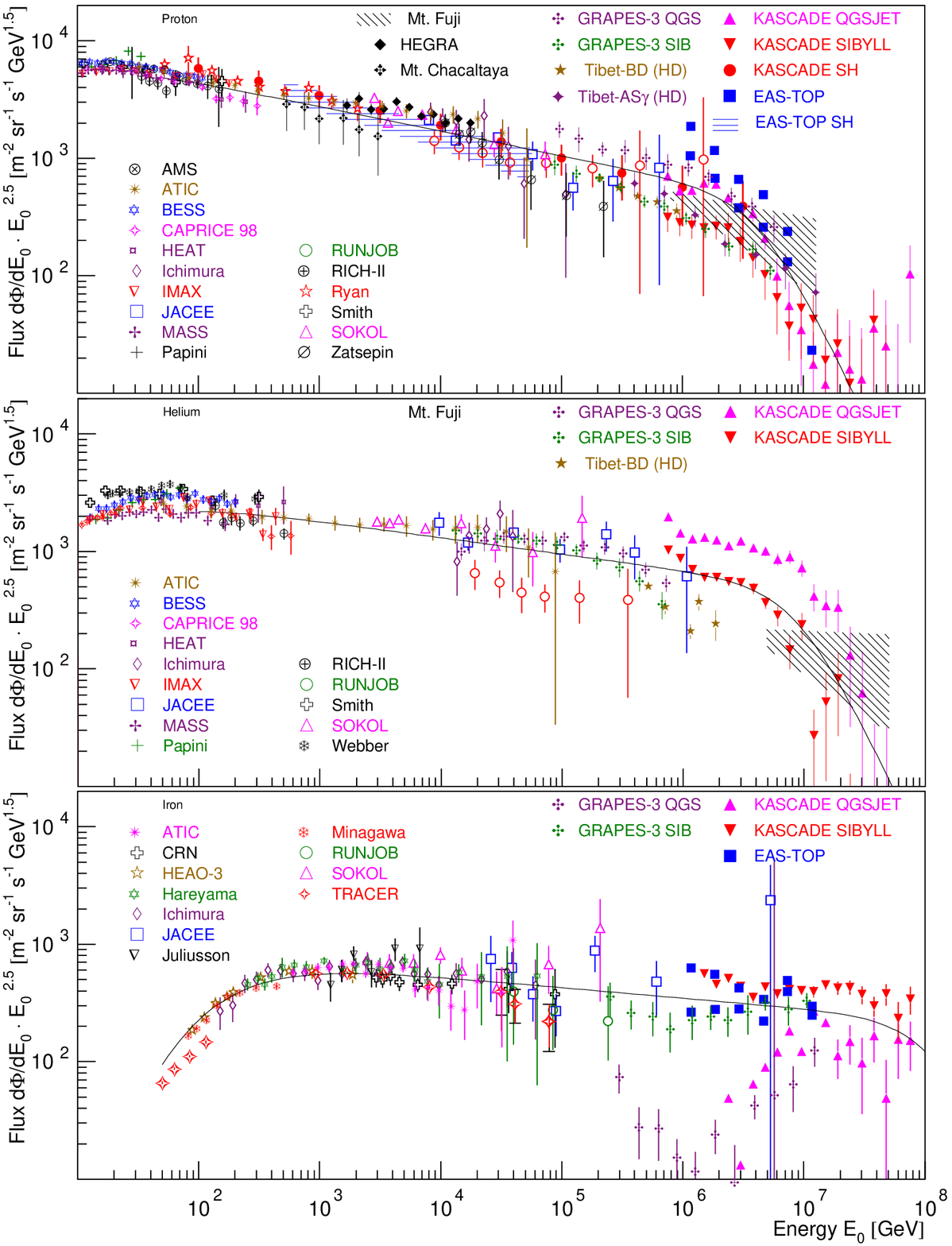,height=27pc}
}
\caption{Compilation of spectral  data $10^{10} - 10^{17}$ eV for  p, He, Fe,
combining balloon, satellite and ground-based measurements. 
From \cite{2005astro.ph..8014H}  
and G. H\"orandel, private communication.
}
\label{pHeFe_Hoerandel}
\end{figure}

\begin{figure}
\centerline{
\psfig{figure=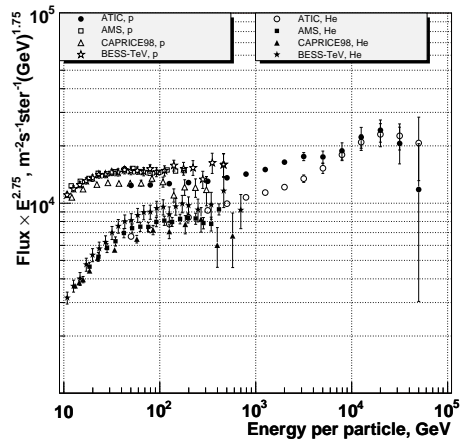,height=15pc} 
}
\caption{Preliminary  spectra of p, He   from ATIC-2, compared with AMS01, BESS-TeV and CAPRICE98. Plot from ATIC collaboration
\cite{Panov2006}. 
 The ATIC-2 data indicate a slighly harder spectrum for He above 1 TeV.
  }
\label{atic2}
\end{figure}

\begin{figure}
\centerline{
\psfig{figure=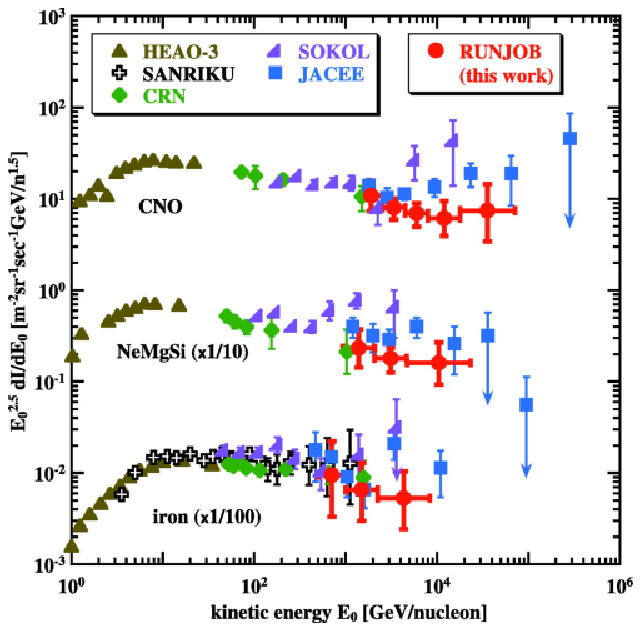,height=15pc}
}
\centerline{
\psfig{figure=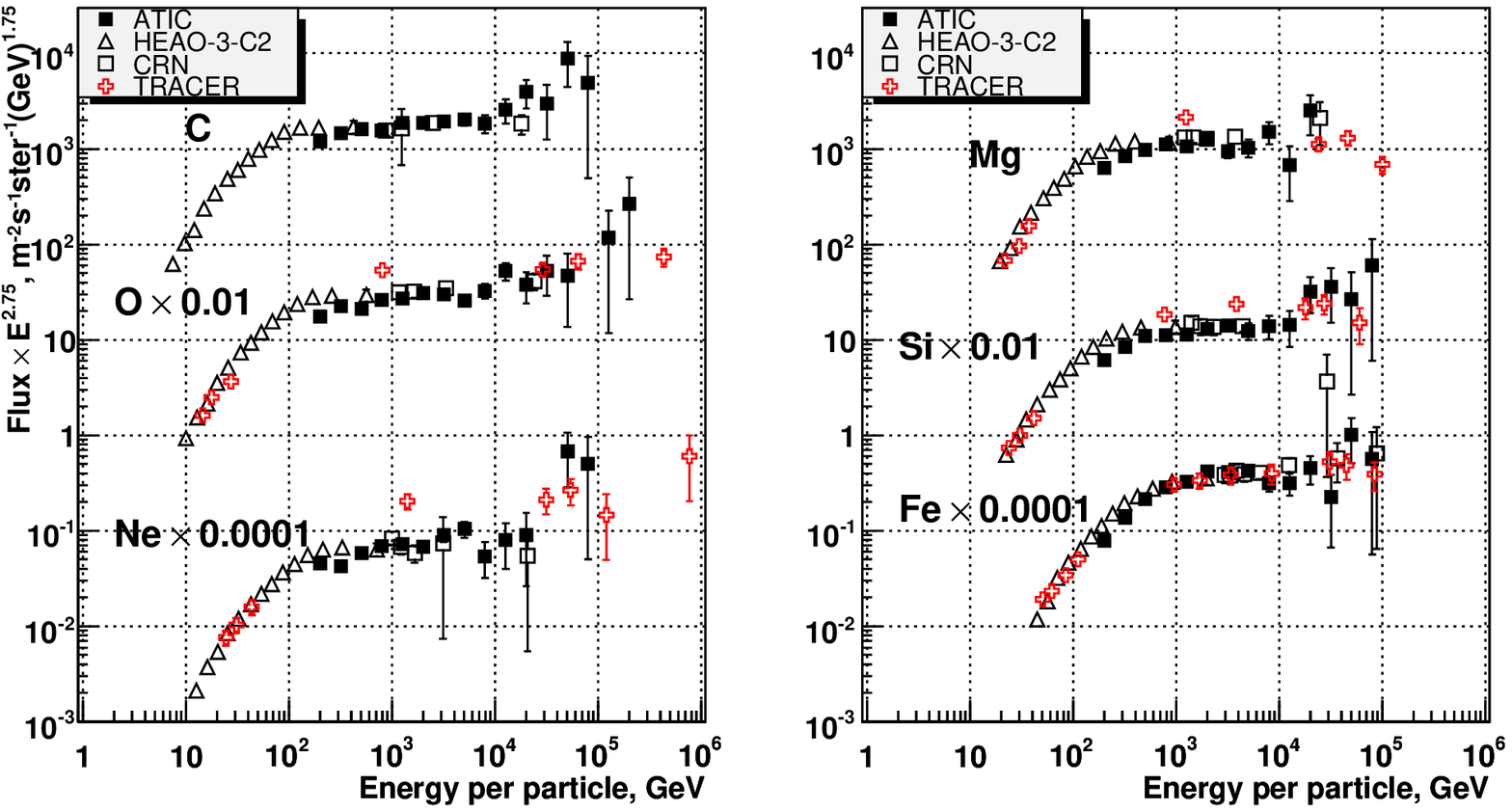,height=15pc}
}
\caption{Compilation of spectral  data for element groups  CNO, NeMgSi, Fe  
\cite{2005ApJ...628L..41D}  
from HEAO-3, SANRIKU, CRN, SOKOL, JACEE and RUNJOB
(upper)
and  of separate even-Z elements from  preliminary ATIC-2, HEAO-3, CRN and TRACER (lower). Plot from ATIC collaboration
\cite{Panov2006}.
The ATIC-2 data suggest a hardening above 10 TeV.
 }
\label{CNO_NeMgSi_Fe_Derbina}
\end{figure}

\begin{figure}
\centerline{
\psfig{figure=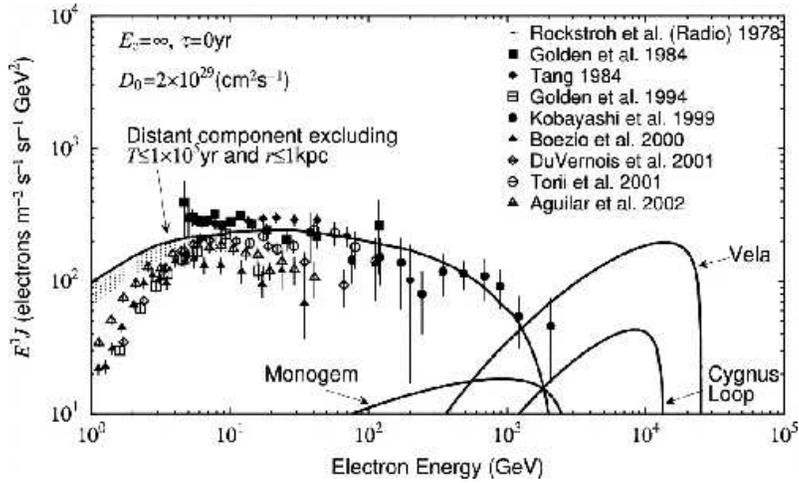,height=15pc}
}
\caption{Measurements of the electron spectrum, including AMS01, CAPRICE94, HEAT and SANRIKU, compared with
possible contributions of distant sources and local supernova remnants,  from  
\cite{2004ApJ...601..340K}. 
 }
\label{electrons_Kobayashi}
\end{figure}

\begin{figure}
\centerline
{\psfig{figure=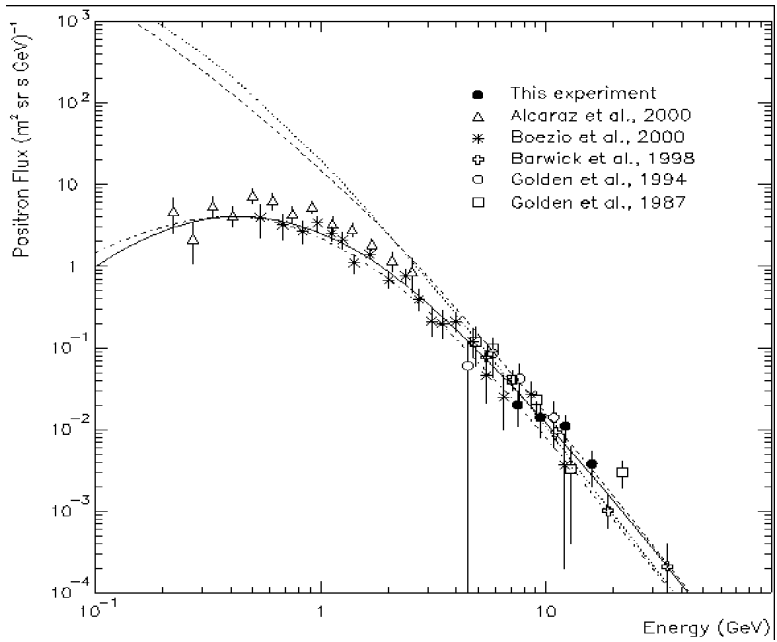,height=20pc}
 }
\caption{Measurements of the positron spectrum,
 including data from MASS91, AMS01, CAPRICE94 and  HEAT, 
  from
\cite{2002A&A...392..287G}.  
Propagation calculations for interstellar (upper curves) and modulated (lower curves) are shown.
Dotted, dot-dashed: GALPROP \cite{1998ApJ...493..694M}; 
dashed, solid: \cite{2001AdSpR..27..687S}. 
}
\label{positrons_Grimani_Boezio}
\end{figure}

\begin{figure}
\centerline
{\psfig{figure=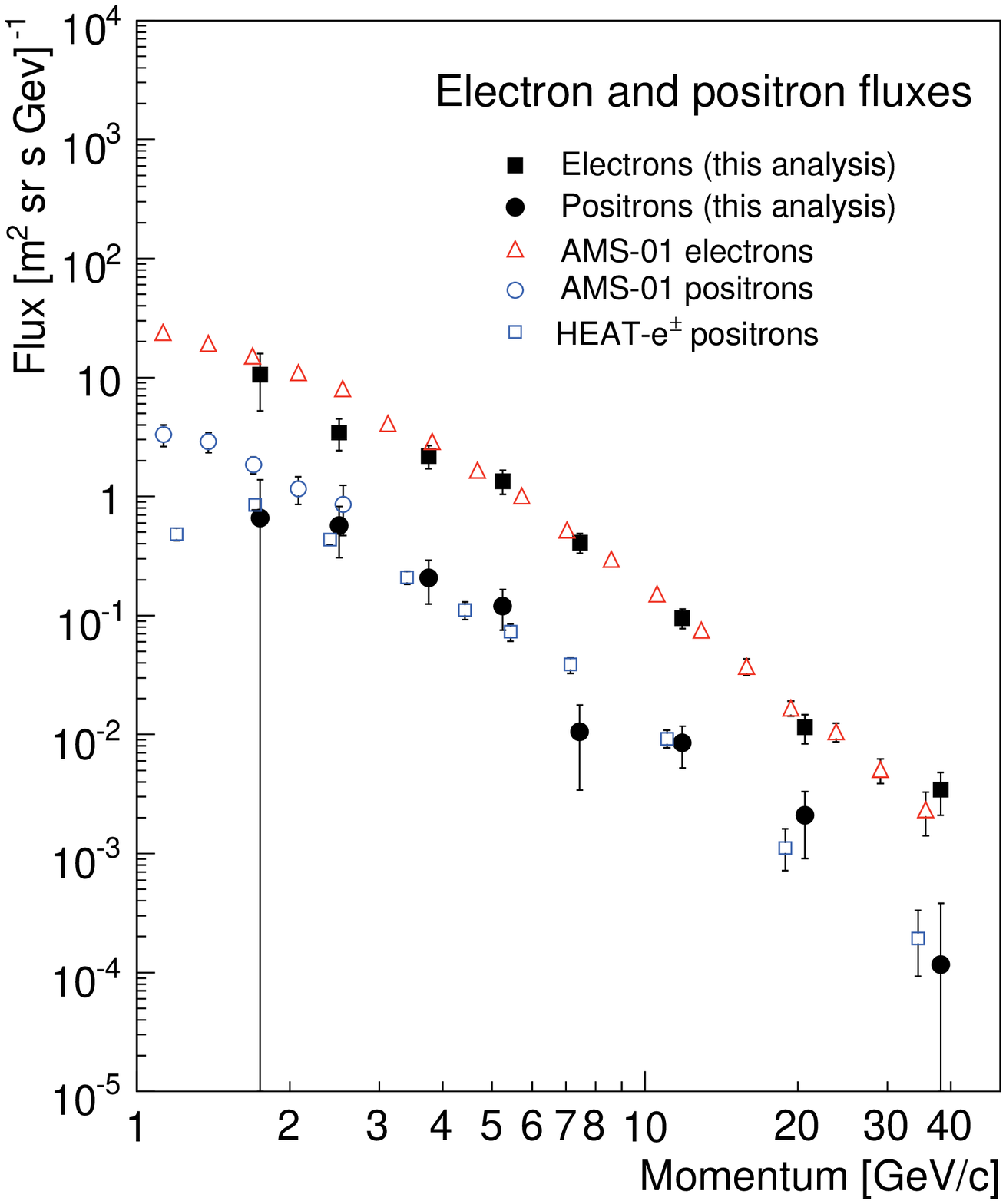,height=15pc}
\psfig{figure=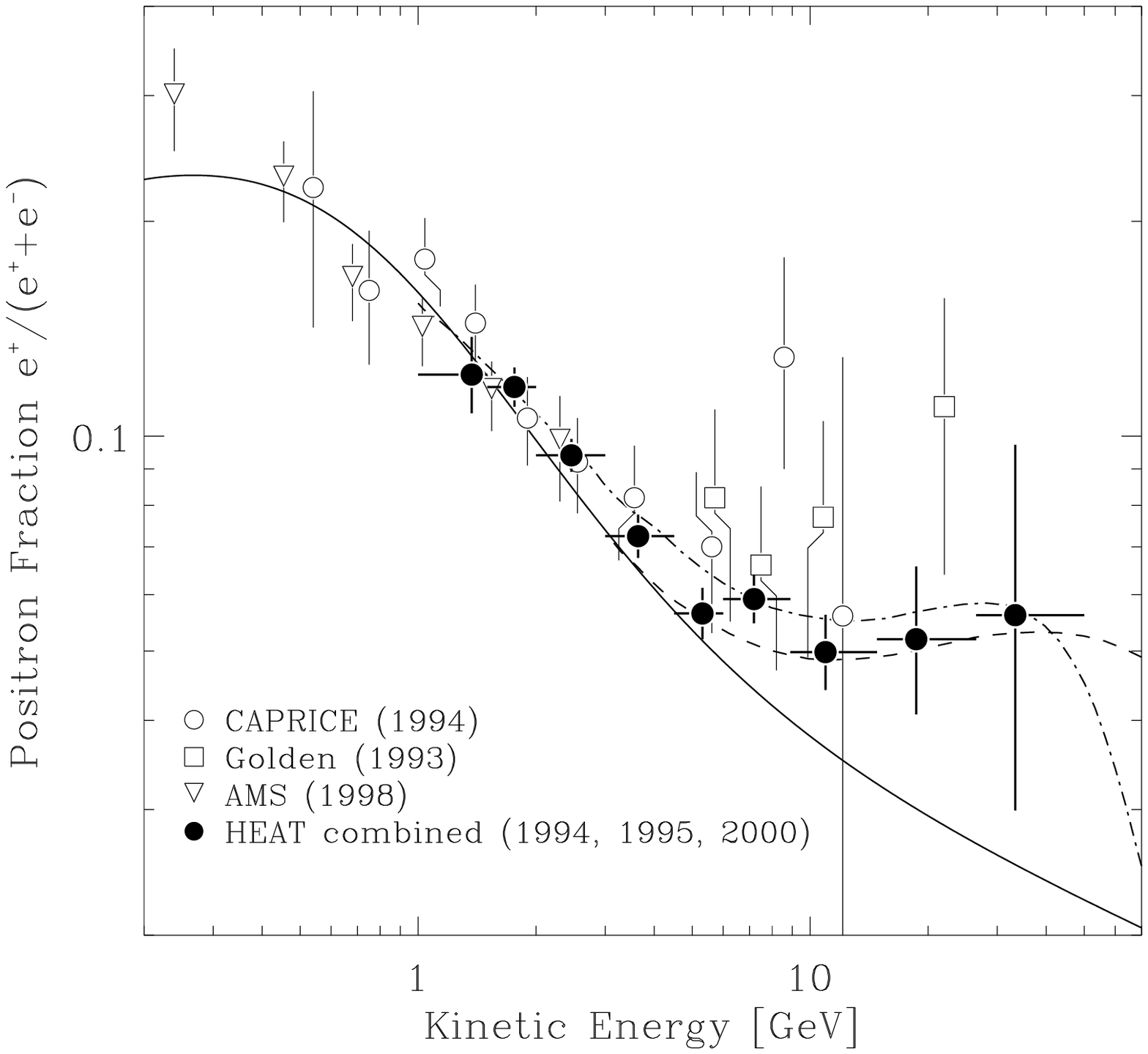,height=15pc}
 }
\caption{Left: Measurements of the electron and  positron spectrum  by AMS01 and HEAT, from 
\cite{2006astro.ph..5254G}. 
Right:
$ e^+/(e^+ + e^-)$ ratio including HEAT and  AMS01 
\cite{2004PhRvL..93x1102B}. 
For the ratio a propagation calculation 
\cite{1998ApJ...493..694M} 
is shown by the solid line, and possible contributions from pulsars (dashed) and dark matter annihiliations (dash-dot).
 }
\label{electrons_AMS_Gast}
\end{figure}

\begin{figure}
\centerline
{\psfig{figure=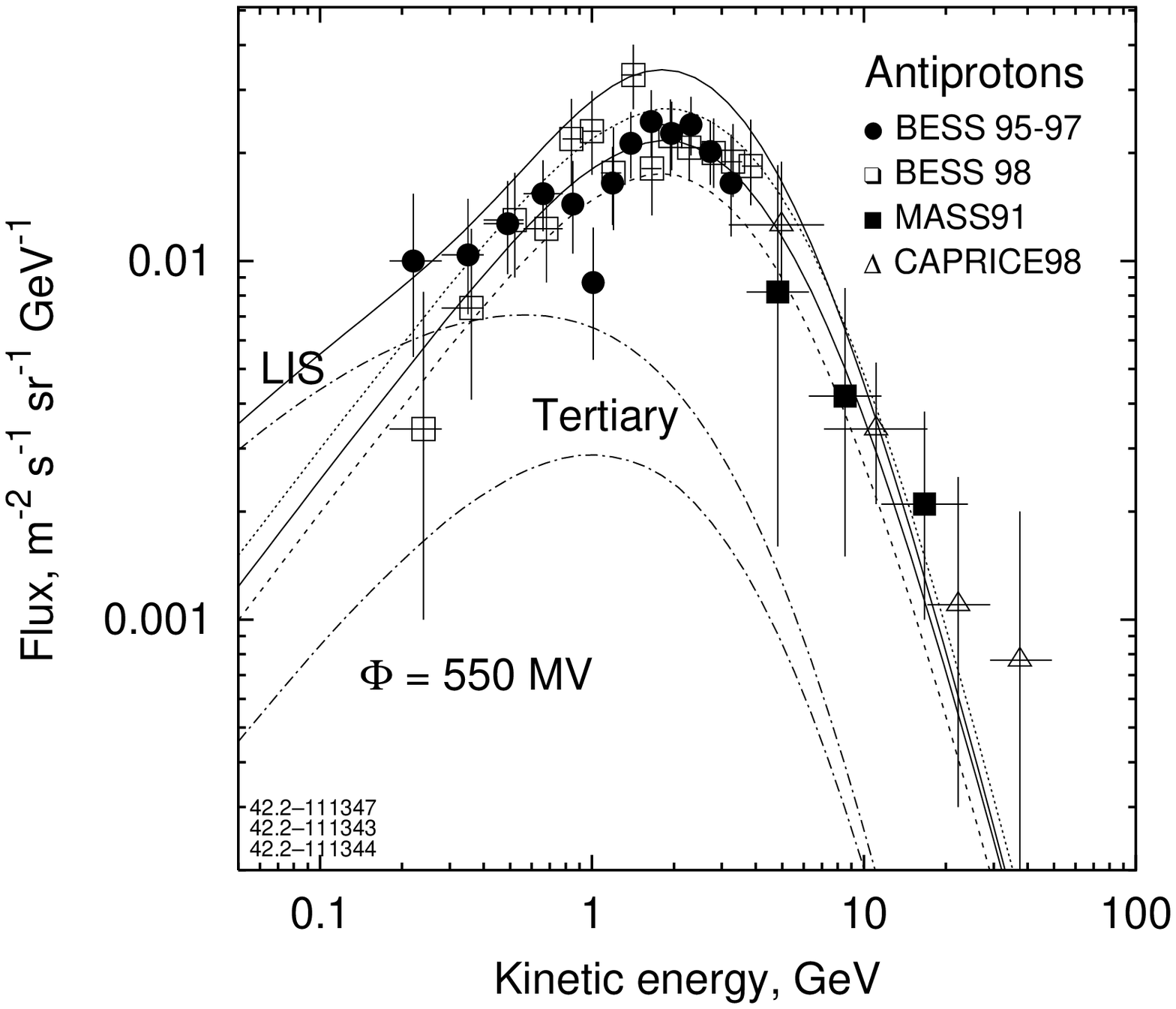,height=15pc}
 }
\caption{Measurements of the antiproton spectrum from MASS91, CAPRICE98 and BESS, compared to a  propagation calculation, from  
\cite{2005AdSpR..35..156M}. 
The dashed and dotted lines illustrate the
sensitivity of the calculated (modulated) antiproton flux to the
normalization of the diffusion coefficient.
LIS marks the local interstellar spectrum. 
 }
\label{antiprotons_Moskalenko}
\end{figure}

\section{Cosmic-ray Propagation: Theory} 

\subsection{Basics and Approaches}

We present the basic concepts of CR  propagation, and techniques for  relating these to observational data. 
Practically all our  knowledge of CR propagation  comes via secondary CR, with additional information from \grays and synchrotron radiation.
It is useful at the outset to point out why secondary nuclei in particular are a good probe of CR propagation:
the fact that the primary nuclei are measured (at least locally) means that the
secondary production functions can be computed from primary spectra, cross-sections and interstellar gas densities
with reasonable precision;
the secondaries can then be `propagated' and compared with observations.

Since the realization that CR fill the Galaxy it has been clear that nuclear interactions imply that 
their composition contains information on their propagation
\cite{1950PhRv...80..943B}. 
A historical event was the  arrival of satellite measurements of isotopic Li, Be, B in the 1970's
\citep{1975ApJ...201L.145G}. 
Since then the subject has expanded enormously with models of increasing degrees of sophistication.
The simple observation that the observed composition of CR is different from that of solar,
in that rare solar-system nuclei like Boron are abundant in CR, proves the importance of propagation in the interstellar medium.
The canonical `few g cm$^{-2}$' of traversed material is one of the widest-known facts of cosmic-ray physics.

 At present we believe that the diffusion model with possible inclusion of convection provides the
 most adequate description of CR transport in the Galaxy at energies below about $10^{17}$ eV so we begin by presenting this model.



\subsection{Propagation equation}


\newcommand{\Dpp}{D_{pp}}
\newcommand{\Dxx}{D_{xx}}
\newcommand{\ddp}{{\partial\over\partial p}}

The CR propagation equation for a particular particle species can be written in the general form:
\begin{eqnarray}
\label{A.1}
{\partial \psi (\vec r,p,t) \over \partial t} 
&= &
  q(\vec r, p, t)                                             
   + \vec\nabla \cdot ( \Dxx\vec\nabla\psi - \vec V\psi )   \nonumber  \\
&   +& \ddp\, p^2 \Dpp \ddp\, {1\over p^2}\, \psi                  
   - {\partial\over\partial p} \left[\dot{p} \psi
   - {p\over 3} \, (\vec\nabla \cdot \vec V )\psi\right]
   - {1\over\tau_f}\psi - {1\over\tau_r}\psi\ 
\end{eqnarray}

\noindent
where $\psi (\vec r,p,t)$ is the CR density per unit of total
particle momentum $p$ at position $\vec r$, $\psi(p)dp = 4\pi p^2 f(\vec p)dp $ in terms of
phase-space density $f(\vec p)$, $q(\vec r, p)$ is the source term including primary, spallation and decay contributions,
$\Dxx$ is the spatial diffusion coefficient, $\vec V$ is the convection
velocity,
 diffusive reacceleration is described as diffusion in momentum space
and is determined by the coefficient $\Dpp$\ , $\dot{p}\equiv dp/dt$
is the momentum gain or loss rate, $\tau_f$ is the time scale for loss by
fragmentation, and $\tau_r$ is the time scale for  radioactive
decay.

CR sources are usually assumed
to be concentrated near the Galactic disk and to have a  radial distribution  like for example supernova remnants (SNR).
A source injection spectrum and its isotopic composition are required; composition is  usually initially based on primordial solar
but can be determined iteratively from the CR data themselves for later comparison with solar.
The spallation part of $q(\vec r, p,t ) $ depends on all progenitor species and their energy-dependent cross-sections,
 and the gas density $n(\vec r)$; it is generally assumed that the spallation products have the same kinetic energy per nucleon as the progenitor.
K-electron capture  and electron stripping can be included  via  $\tau_f$ and $q$.
 $\Dxx$   is in general a function of  $(\vec r, \beta, p/Z)$ where $\beta=v/c$ and $Z$ is the charge, and $p/Z$ determines the gyroradius in a given magnetic field;
 $\Dxx$ may be  isotropic, or more realistically anisotropic,
and may be influenced by the CR themselves (e.g. in  wave-damping  models).
 $\Dpp$\ is related to  $\Dxx$ by  $\Dpp\Dxx\propto p^2$, with the proportionality constant depending
 on the  theory of stochastic reacceleration
\cite{1990acr..book.....B,1994ApJ...431..705S} 
 as described in Section 2.5.
 $\vec V$ is a function of   $\vec r$ and depends on the nature of the Galactic wind.
The term in $\vec\nabla \cdot \vec V$ represents  adiabatic momentum  gain or loss in the
 non-uniform flow of gas with a frozen-in magnetic field whose inhomogeneities scatter the CR.
$\tau_f$ depends on the total spallation cross-section and    $n(\vec r)$.
$n(\vec r)$ can be based on surveys of atomic and molecular gas, but can also incorporate 
small-scale variations such as the region of low gas density  surrounding the Sun.
The presence of interstellar helium  at about 10\% of hydrogen by number must be included;
heavier components of the ISM are not important for producing CR by spallation.
This equation only treats continous momentum-loss; catastrophic losses can be included via  $\tau_f$ and $q$.
CR electrons, positrons and antiprotons propagation constitute just special cases of this equation, differing only 
in their energy losses and production rates.

The {\it boundary conditions} depend on the model; often $\psi=0$ is assumed at the `halo boundary' 
 where particles escape into intergalactic space, but
this obviously just an approximation (since the intergalactic flux is not zero) which can be relaxed for  models
with a physical treatment of the boundary.

Equation (1) is a time-dependent equation; usually the steady-state solution is required, which can be
obtained either by setting $\partial \psi / \partial t = 0$ or following the time dependence
until a steady state is reached; the latter procedure is much easier to implement numerically.
The time-dependence of $q$ is neglected unless effects of nearby recent sources  or the stochastic nature of sources are being studied.
Starting with the solution for the heaviest primaries and using this to compute the spallation source
for their products, the complete system can be solved including secondaries, tertiaries etc.
Then the CR spectra at the solar position can be compared with direct observations, including solar modulation if required.

Source abundances are determined iteratively, comparing propagation calculations with data;
for nuclei with very small source abundances, the source values are masked by secondaries and cross-section uncertainties  and are therefore hard to determine.
Webber 
\citep{1998SSRv...86..239W} 
 gives a ranking from `easy' to `impossible' for the possibility of getting the source abundances using ACE data.
A recent review of the high-precision abundances from ACE  is in
\citep{2001SSRv...99...15W} 
and for Ulysses in
\citep{2001SSRv...99...41C}. 
For a useful summary of the various astrophysical abundances relevant to interpreting CR abundances see 
 \citep{2005ApJ...634..351B}. 

\subsection{Diffusion} 

The concept of CR diffusion explains why energetic charged particles
have highly isotropic distributions and why they are retained well in the
Galaxy. The Galactic magnetic field which tangles the trajectories of
particles plays a crucial role in this process.  Typical values of the
diffusion coefficient found from  fitting to CR  data is $\Dxx\sim(3 - 5)\times10^{28}$ cm$^{2}$ s$^{-1}$ at energy 
$\sim$1~GeV/n and it increases with magnetic rigidity as $R^{0.3}-R^{0.6}$ in different
versions of the empirical diffusion model of CR propagation.

On the ``microscopic level'' the diffusion of CR  results from 
particle scattering on random MHD waves and discontinuities. The effective
``collision integral'' for charged energetic particles moving in a magnetic
field with small random fluctuations $\delta B\ll B$ can be taken from the
standard quasi-linear theory of plasma turbulence \cite{Kennel66}.
The wave-particle interaction is of resonant character so that an energetic
particle is predominantly scattered by those irregularities of magnetic field
which have their projection of the wave vector on the average magnetic field
direction equal to $k_{\parallel}=\pm s/\left(  r_{\mathrm{g}}\mu\right)  $,
where $\mu$ is the particle pitch angle. The integers $s=0,1,2...$ correspond
to  cyclotron resonances of different orders. The efficiency of scattering
depends on the polarization of the waves and on their distribution in
$\mathbf{k}$-space. The first-order resonance $s=1$\ is the most important for
the isotropic and also for the one-dimensional distribution of random MHD
waves along the average magnetic field. In some cases -- for calculation of
scattering at small $\mu$ and for calculation of perpendicular diffusion --
the broadening of resonances and magnetic mirroring effects should be taken
into account. The resulting spatial diffusion is strongly anisotropic locally
and goes predominantly along the magnetic field lines. However, strong
fluctuations of magnetic field on  large scales $L\sim100$ pc, where the
strength of the random field is several times higher than the average field
strength, lead to the isotropization of global CR diffusion in the
Galaxy. The rigorous treatment of this effect is not trivial, since the
field is almost static and the strictly one-dimensional diffusion along the
magnetic field lines does not lead to non-zero diffusion perpendicular to
$\mathbf{B}$, see \cite{2002PhRvD..65b3002C}
%
%
and the references cited there.

Following several detailed reviews of the theory of CR  diffusion
\cite{1971RvGSP...9...27J,Toptygin85,1990acr..book.....B,2002cra..book.....S} 
%
%
%
the diffusion coefficient  at $r_{\mathrm{g}}<L$\ can be roughly
estimated as $\Dxx\approx\left(  \delta B_{\mathrm{res}}/B\right)  ^{-2}%
vr_{\mathrm{g}}/3$, where $\delta B_{\mathrm{res}}$ is the amplitude of random
field at the resonant wave number $k_{\mathrm{res}}=1/r_{g}$. The spectral
energy density of interstellar turbulence has a power law form
$w(k)dk\sim k^{-2+a}dk$, $a=1/3$ over a wide range of wave numbers
$1/(10^{20}$ \textrm{cm}$)<k<1/(10^{8}$ \textrm{cm}$)$, see
\cite{2004ARA&A..42..211E},
%
and the strength of the random field at the main scale is $\delta B\approx5$ $\mu
$G. This gives an estimate of the diffusion coefficient $\Dxx\approx$
$2\times10^{27}\beta R_{_{\mathrm{GV}}}^{1/3}$ cm$^{2}$ s$^{-1}$ for all CR
particles with magnetic rigidities $R<10^{8}$ GV, in a fair agreement with the
empirical diffusion model (the version with distributed reacceleration). The
scaling law $\Dxx\sim R^{1/3}$ is determined by the value of the exponent
$a=1/3$, typical for a Kolmogorov spectrum. Theoretically
\cite{1995ApJ...438..763G} 
 the
Kolmogorov type spectrum might refer only to some part of the MHD turbulence
which  includes the (Alfvenic) structures strongly elongated along the
 magnetic-field direction and which are not able to provide the significant scattering
and required diffusion of cosmic rays. In parallel, the more isotropic (fast
magnetosonic) part of the turbulence, with a smaller value of random field at
the main scale and with the exponent $a=1/2$ typical for the Kraichnan type turbulence
spectrum, may exist in the interstellar medium 
\cite{2004ApJ...614..757Y}. 
 The Kraichnan spectrum
gives a scaling $\Dxx\sim R^{1/2}$ which  is close to the high-energy asymptotic form
of the diffusion coefficient obtained in the `plain diffusion' version of
the empirical propagation model. Thus the approach based on  kinetic theory
gives a proper estimate of the diffusion coefficient and predicts a power-law
dependence of diffusion on magnetic rigidity, but the  determination of the actual
diffusion coefficient has to  be done with the help of  empirical
models of CR  propagation in the Galaxy.

\subsection{Convection}

While the most frequently considered mode of CR transport  is diffusion, 
the existence of galactic  winds in many galaxies suggests that convective (or advective) transport
could be important.
Winds are common in galaxies and can be CR driven \cite{2000Ap&SS.272....3B}.  
 CR play a dynamical role in galactic halos
\citep{1991A&A...245...79B,1993A&A...269...54B}. 
Convection not only transports CR, it can also produce adiabatic energy losses as the wind speed increases away from the disk.
Convection was first considered by 
\cite{1976ApJ...208..900J} 
and followed up by
\cite{1977ApJ...215..677O,1977ApJ...215..685O,1978ApJ...222.1097J,1979ApJ...229..747J,1993A&A...267..372B}. 
Both  1-zone and 2-zone models have been studied: 
a 1-zone model has convection and diffusion everywhere, a 2-zone model  has diffusion alone up to some distance from the plane, and diffusion plus convection beyond.

A   recent AnnRev on galactic winds 
\citep{2005ARA&A..43..769V} 
does not mention CR, surprisingly.
In the same volume, Cox's article on the Three Phase Interstellar Medium
\citep{2005ARA&A..43..337C} 
 includes CR as a basic component.
Direct evidence for winds in our own Galaxy seems to be confined to the Galactic centre
region from X-ray images.
 However Cox
\citep{2005ARA&A..43..337C} 
 is not sure there is a wind:
`A Galactic wind may be occurring, but I do not believe that it carries off a significant fraction of the supernova power from the Solar Neighborhood because it would carry off a similar power in the pervading cosmic rays...'

For {\it one-zone} diffusion/convection models
a good diagnostic  is the energy-dependence of the secondary-to-primary ratio: 
a purely convective transport would have no energy dependence (apart from the velocity-dependence of the reaction rate), contrary to what is observed.
If the diffusion rate decreases with decreasing energy,  any convection will eventually take over
and cause the secondary-to-primary ratio to flatten at low energy: this is observed but 
convection 
( proposed 
\cite{1979ApJ...229..747J} 
  to explain just this effect)
 does not reproduce e.g. B/C very well
\cite{1998ApJ...509..212S}. 
Another test is provided by radioactive isotopes which effectively constrain the wind speed to $<$$10$ km s$^{-1}$ kpc$^{-1}$
for a speed increasing linearly with distance from the disk
\cite{1998ApJ...509..212S}. 
A value of   $\approx 15$ km s$^{-1}$ (constant speed wind) is required to fit B/C 
even in the presence of reacceleration
according to
\citep{2002A&A...394.1039M} 
which can be compared to  $30$ km s$^{-1}$   in the wind model of
\citep{2001ApJ...547..264J}; 
the latter value implies an energy dependence of the  diffusion  coefficient which may conflict with CR anisotropy.

\citep{1997A&A...321..434P} 
studied a self-consistent  {\it two-zone} model with a wind driven by CR and thermal gas in a rotating Galaxy.
The CR propagation is entirely diffusive in a zone $|z|< 1$ kpc, and diffusive-convective outside.
CR reaching the convective zone do not return, so it acts as a halo boundary with  height varying with energy and Galactocentric radius.
It is possible to explain the energy-dependence of the secondary-to-primary ratio with this model,
and it is also claimed to be consistent with radioactive isotopes.
The effect of a Galactic wind on the radial CR gradient has been investigated 
\citep{2002A&A...385..216B}; 
they constructed a self-consistent model with the wind driven by CR, and with anisotropic diffusion.
The convective velocities involved in the outer zone are large (100 km s$^{-1}$) but this model is still
consistent with radioactive CR nuclei which set a much lower limit
\citep{1998ApJ...509..212S}, 
since this limit is only applicable in the inner zone. 
Observational support of such models would require direct evidence for a Galactic wind in the halo.


\subsection{Reacceleration}

In addition to spatial diffusion, the scattering of CR particles
on randomly moving MHD waves leads to stochastic
acceleration which is described in the transport equation as diffusion in
momentum space with some diffusion coefficient $D_{pp}$. One can estimate it
as $D_{pp}=p^{2}V_{\mathrm{a}}^{2}/\left(  9\Dxx\right)  $ where the Alfv\'en
velocity $V_{\mathrm{a}}$ is introduced as a characteristic velocity of
weak disturbances propagating in a magnetic field, see
\citep{1990acr..book.....B,2002cra..book.....S}
%
%
for rigorous formulas.

    Distributed acceleration in the entire Galactic volume cannot serve as
the main mechanism of acceleration of  CR at least in the
energy range $1-100$ GeV/n.  In this case the particles of higher
energy would spend longer in the system, which would result
in an increase of the relative abundance of secondary nuclei as energy increases, contrary to observation. This
argument does not hold at low energies where  distributed acceleration may
be strong and it may explain the existence of peaks in the ratios of
secondary to primary nuclei at about $1$ GeV/n if the
distributed acceleration becomes
significant at this energy. The process of distributed acceleration in the
interstellar medium is  also referred to as 
`reacceleration' to distinguish it from the primary acceleration process
which occurs in the CR sources.
It has been shown
\cite{1986ApJ...300...32S,1994ApJ...431..705S} 
that the observed dependence of abundance of secondary nuclei on energy can
be explained in the model with reacceleration if the CR
 diffusion coefficient varies as
a single power law of rigidity $\Dxx\sim R^{a}$ with an exponent $a\sim
0.3$ over the whole energy range (corresponding to  particle scattering
on MHD turbulence with a Kolmogorov spectrum), and if  the Alfv\'en velocity is
$V_{\mathrm{a}} \sim30$ km s$^{-1}$, which is close to its actual value in the
interstellar medium. 

In addition to stable secondary nuclei, the secondary K-capture isotopes are
useful for the study of possible reacceleration in the interstellar medium
\cite{1985Ap&SS.114..365L}. 
The isotopes $^{37}$Ar, $^{44}$Ti, $^{49}$V, $^{51}$Cr and some others
 decay rapidly by electron capture at low energies where energetic ions can have
an orbital electron. The probability to have an orbital electron  depends strongly
 on energy and because of this the abundance of these isotopes and of
their decay products are strong functions of energy and sensitive to  changes of
particle energy in the interstellar medium. The first measurements of an
energy-dependent decay of $^{49}$V and $^{51}$Cr in CR
\cite{2000AIPC..528..406N} 
were used to test the rate of distributed interstellar reacceleration
\cite{2001AdSpR..27..737J} 
but  refinement of nuclear
production cross sections is required to draw  definite conclusions.

The gain of particle energy in the process of reacceleration is accompanied by a
corresponding energy loss of the interstellar MHD turbulence. According to
calculations
\citep{2006ApJ...642..902P}
%
the dissipation on CR          may significantly influence the Kraichnan nonlinear
cascade of waves at less $10^{13}$ cm and even terminate the cascade
at small scales. This results in a
self-consistent change of the rigidity dependence of diffusion coefficient with
a steep rise of $\Dxx$ to small rigidities.
The scheme explains the high-energy scaling of diffusion $\Dxx\sim R^{0.5}$
and offers an explanation of the observed energy dependence of
primary to secondary ratios.

As mentioned above, the data on secondary nuclei provide evidence against strong
reacceleration in the entire Galaxy at $1-100$ GeV/n. However,
the spectra of secondaries can be considerably modified due to  processes 
in the source regions, with a small  total Galactic  volume filling factor,  for the regions where the high-velocity
SNR shocks accelerate primary CR. Two effects could be operating there and both lead to the
production of a component of secondaries with flat energy spectra
\cite{1987ApJ...316..676W,2003A&A...410..189B}. 
One effect is the production of secondaries in SNR   by the spallation of
primary nuclei which have a flat source energy spectrum close to
$E^{-2}$. Another effect is the direct acceleration by strong SNR shocks of background secondary
nuclei residing in the interstellar medium; again the
secondaries acquire the flat source energy spectrum.  Calculations 
%
%
\cite{2003A&A...410..189B}
%
showed that these effects might produce a flat component of secondary
nuclei rising above   the standard steep spectrum of
secondaries at energies above about $100$ GeV/n.


\subsection{Galactic structure}

Almost all aspects of Galactic structure affect CR propagation,
but the most important are the gas content for secondary production 
 and the interstellar radiation field and magnetic field for electron energy losses.
The magnetic field is clearly also important for diffusion but the precise absolute magnitude and large-scale structure
are less important (at least for CR below $10^{15}$ eV) than the turbulence properties.

The distribution of atomic hydrogen is reasonably well known from 21-cm surveys,
but the molecular hydrogen is less well known since it has to use the CO molecular tracer
and the conversion factor is hard to determine and may depend on position in the Galaxy.
In fact CR--gas interactions  provide one of the best methods to determine the molecular hydrogen content of the Galaxy, 
because of its basic simplicity,  as we describe in the chapter on gamma rays.
For more details we refer to another review \cite{2004cgrs.conf..279M}. 

The Galactic magnetic field can be determined from pulsar rotation and dispersion  measures combined with a model for the distribution of ionized gas.
A large-scale field of a few $\mu G$  aligned with spiral arms exists, but there is no general agreement on the details
\cite{2001SSRv...99..243B}. 
One recent analysis gives a bisymmetric model for the large-scale Galactic magnetic field with reversals on arm-interarm boundaries 
\cite{2006ApJ...642..868H}. 
Independent estimates of the strength and distribution of the field can be made by simultaneous analysis of radio synchrotron, CR and \gray data,
and these confirm a value of a few $\mu G$, increasing towards the inner Galaxy
\cite{2000ApJ...537..763S}. 
A lot of effort has gone into constructing magnetic  field models to study propagation at energies $>$$10^{15}$ eV  where
the Larmor radius is large enough for the global topology to be important; this is relevant to CR anisotropy and the search for point sources.
Since this is excluded from our review we refer the reader to 
 \cite{2002ApJ...572..185A,2006ApJ...639..803T,2005astro.ph.10444K}. 

The interstellar radiation field (ISRF) comes from stars of all types and is processed by absorption and re-emission by interstellar dust;
it extends from the far-infrared though optical to the UV. Computing the ISRF is difficult, but a great deal
of new information on the stellar content of the Galaxy and dust is now available to make better models
for use in propagation codes
\cite{2006ApJ...640L.155M}. 

The local environment around the Sun
\cite{2006astro.ph..6743F} 
 is also important, for example the local bubble can have an effect on radioactive nuclei as described in Sec 3.2.
Extensive coverage of the local environment including CR is in
 \cite{2006sjsg.book.....F}. 

\subsection{Interactions} 
This large subject is well covered in the literature.
Details of the essential processes with references have been conveniently collected in our series of papers:
energy losses of nuclei and electrons
 \cite{1998ApJ...509..212S}, 
bremsstrahlung and synchrotron emission
\cite{2000ApJ...537..763S}, 
inverse-Compton emission including anisotropic scattering
\cite{2000ApJ...528..357M}, 
pion production of $\gamma$-rays, electrons and positrons
\cite{1998ApJ...493..694M}. 
Pion production has recently been studied in great detail using modern particle-physics codes
\cite{2006ApJ...647..692K,2006PhRvD..74c4018K},  
the former giving spectra harder by 0.05 in the index and \gray yields somewhat higher at a few GeV than  
older treatments. New more accurate parameterizations will  be important for the new generation of CR and \gray experiments.
A useful guide to spallation cross-section measurements and models is  the
contribution by J. Connell in
\cite{2001SSRv...99..353M}, 
and a summary of recent advances is in 
\cite{2004AdSpR..34.1288M,2005AIPC..769.1612M,2006AdSpR..38.1558D}. 
 Accounts of radioactive and K-capture processes are in
\citep{1978NuPhA.310....1E,1998PhRvL..81..281M,1998PhRvL..80.2085W,2000AIPC..528..406N,2001AdSpR..27..717S}. 


\subsection{Weighted Slabs and Leaky Boxes} 

As mentioned at the start of this section,
 at present we believe that the diffusion model with possible inclusion of convection provides the most adequate description of CR transport in the Galaxy at energies below about $10^{17}$ eV. The closely related  leaky-box and weighted slab formalisms have provided the basis for most of the literature interpreting CR data.

In the leaky-box model, the diffusion and convection terms are approximated by the leakage term with some characteristic escape time of CR from the Galaxy. The escape time
 $\tau_{esc}$ may be a function of particle energy (momentum), charge, and mass number if needed, but it does not depend on the spatial coordinates. There are two cases when the leaky box equations can be obtained as a correct approximation to the diffusion model: 1) the model with fast CR diffusion in the Galaxy and particle reflection at the CR halo boundaries with some probability to escape    
\cite{1964ocr..book.....G}, 
2) the formulae for CR         density in the Galactic disk in the flat halo model $(z_h \ll R)$ with thin source and gas disks ($z_{gas} \ll z_h)$ 
which are formally equivalent to the leaky-box model formulae in the case when stable nuclei are considered
\citep{1976RvMP...48..161G}. 
 The nuclear fragmentation is actually determined not by the escape time $\tau_
{esc}$ but rather by the escape length in g cm$^{-2}$: $x = v\rho\tau_{esc}$ ,
where $\rho$ is the average gas density of interstellar gas in a galaxy with the volume of the cosmic ray halo included.

The solution of a system of coupled transport equations for all isotopes involved in the process of nuclear fragmentation is required for studying CR propagation. A powerful method, the weighted-slab technique, which consists of splitting  the problem into astrophysical and nuclear parts was suggested for this problem
\cite{1960ICRC....3..220D,1964ocr..book.....G} 
 before the modern computer epoch. The nuclear fragmentation problem is solved in terms of the slab model wherein the CR beam is allowed to traverse a thickness $x$ of the interstellar gas and these solutions are integrated over all values of $x$ weighted with a distribution function $G(x)$  derived from an astrophysical propagation model. In its standard realization
\cite{1981ApJ...247..362P,1987ApJS...64..269G} 
 the weighted-slab method breaks down for low energy CR where one has strong energy dependence of nuclear cross sections, strong energy losses, and energy dependent diffusion. Furthermore, if the diffusion coefficient depends on the nuclear species the method has rather significant errors. After some modification
 \citep{1996ApJ...465..972P} 
 the weighted-slab method becomes rigorous for the important special case of separable dependence of the diffusion coefficient on particle energy (or rigidity) and position with no convective transport. The modified weighted-slab method was applied to a few simple diffusion models in
\cite{2001ApJ...547..264J,2001AdSpR..27..737J}. 
 The weighted-slab method can also be applied to the solution of the leaky-box equations. It can  easily be shown that the leaky-box model has an exponential distribution of path lengths $G(x) \propto \exp(-x/X)$ with the mean grammage equal to the escape length $X$.

In a purely empirical approach, one can try to determine the shape of the distribution function $G(x)$ which best fits the data on abundances of stable primary and secondary nuclei 
\cite{1970ARNPS..20..323S}.  
 It has been established that the shape of $G(x)$ is close to  exponential: $G(x) \propto \exp(-x/X(R,\beta))$, and this justifies the use of the leaky-box model in this case.
 There are several recent calculations of  $G(x)$ 
\cite{1998ApJ...505..266S,2000AIPC..528..421D,2001ApJ...547..264J,2001AdSpR..27..737J}.  

The possible existence of truncation, a deficit at small path lengths (below a few g cm$^{-2}$ at energies near 1 GeV/n),
relative to an  exponential path-length distribution,      
    has been discussed for decades 
\cite{1970ARNPS..20..323S,1987ApJS...64..269G,1993ApJ...402..188W,1996A&A...316..555D}.  
The problem was not solved mainly because of cross-sectional uncertainties. In a consistent theory of CR diffusion and nuclear fragmentation in the cloudy interstellar medium, the truncation  occurs naturally if some fraction of CR sources resides inside  dense giant molecular  clouds
\cite{1990AA...237..445P}. 

For radioactive nuclei, the classical approach is to compute the `surviving fraction' which is the ratio of the observed abundance
to that expected in the case of no decay.
Often the result is given in the form of an effective mean gas density, to be compared with the average density in the Galaxy,
 but this density should not be taken at face value.  The surviving fraction can better be related to physical parameters 
\citep{1998A&A...337..859P}. 
None of these methods can face the complexities of propagation of CR electrons and positrons with their large energy and spatially dependent energy losses.


\subsection{Explicit models} 

Finally the mathematical effort required to put the 3-D Galaxy into a 1-D formalism becomes
overwhelming, and it seems better to work in physical space from the beginning: this
 approach is intuitively simple and easy to interpret.
We can call these `explicit solutions.'
The explicit solution approach including secondaries was
pioneered by
\citep{1976RvMP...48..161G}  
 and  applied to newer  data by
\cite{1992ApJ...390...96W,1993A&A...267..372B} 
 with analytical solutions for 2D diffusion-convection models with a cosmic-ray source distribution,
which however had many restrictive approximations to make them tractable (no energy losses, simple gas model).
More recently  a semi-empirical model which is 2D and includes energy-losses and reacceleration
has been developed
\citep{2001ApJ...555..585M,2002A&A...394.1039M}. 
This is  a closed-form solution expressed as a Green's function to be integrated over the sources.
It incorporates a radial CR source distribution, but the gas model is a simple constant density within the disk.
\cite{2004ApJ...609..173T} 
 give an analytical solution for the time-dependent case with a generalized
gas distribution but  now {\it without energy losses}. (This shows again the problem of handling both
gas and energy losses simultaneously in analytical schemes).

A `myriad sources model'
\cite{2003ApJ...582..330H},  
  which is actually
a Green's function method without energy losses, yields similar results to 
\citep{2004ApJ...613..962S} 
for  the diffusion coefficient and   halo size. But applying the no-energy loss case to ACE data is not really justified,
 and  some defects in their formulation have been pointed out
 \cite{2004ApJ...609..173T}. 
A 3D  analytical propagation  method has been developed
\cite{2004ApJ...612..238S,2006ApJ...642..882S} 
 with energy loss and reacceleration, going via a PLD,
 but it cannot   handle ionization losses properly, see  section 2.2 of
\cite{2006ApJ...642..882S}. 
 They have  no spatial boundaries at all
and rather simplified (exponential)  forms for the gas and other distributions.
An approach adapted to fine-scale spatial and temporal variations has been described 
\cite{2005ApJ...619..314B}.  
This uses a Green's function without energy losses or detailed gas model and hence is limited in its application,
but is useful for studying the effect of discrete sources.

The most advanced explicit solution to date is the fully numerical model  described in  the next section.
Even this has limitations in treating some aspects (e.g. when particle trajectories become important at high energies)
so one might ask whether a fully Monte-Carlo approach (as is commonly done for energies $>10^{15}$ eV)
would not be better in the future, given increasing computing power.
This would allow effects like field-line diffusion (important for propagation perpendicular to the Galactic plane)  to be explicitly included.
However it is still challenging: a GeV particle diffusing with a mean free path of 1 pc 
in a Galaxy with 4 kpc halo height
takes  $\sim$$(4000/1)^2 \approx 10^7 $ scatterings to leave the Galaxy,
which would even now need supercomputers to obtain adequate statistics.
 Hence we expect numerical solution of the propagation equations to remain an important approach
for the foreseeable future.


\subsection{GALPROP }

The GALPROP code
\cite{1998ApJ...509..212S} 
  was created with the following aims:
1.  to enable simultaneous predictions of all relevant observations including CR nuclei, electrons and positrons, \grays and synchrotron radiation,
2. to overcome the limitations of analytical and semi-analytical methods, taking advantage of advances in computing power,  as CR, \gray and other data become more accurate,
3.  to incorporate current information on Galactic structure and source distributions,
4. to provide a publicly-available  code as a basis for further expansion.
The first point is the most important, the idea being that all data relate to the same system, the Galaxy, and one cannot for example allow
a model which fits secondary/primary ratios while not fitting \grays or not being compatible with the known interstellar gas distribution.
There are many simultaneous constraints, and to find one model satisfying all of them is a challenge,  which in fact has not been met up to now.
Upcoming missions should benefit\footnote{GALPROP 
has been adopted as the standard for diffuse Galactic \gray emission  for NASA's GLAST \gray observatory, and
is  also made use of  by the AMS, ACE, HEAT and Pamela collaborations.}.

We give a very brief summary of GALPROP; for details we refer the reader to the relevant papers
\cite{1998ApJ...509..212S,1998ApJ...493..694M,2000ApJ...537..763S,2002ApJ...565..280M,2004ApJ...613..962S,2006ApJ...642..902P} 
and a dedicated website\footnote{\tt http://galprop.stanford.edu}.
The propagation equation (1) is solved numerically on a spatial grid, either in 2D with cylindrical symmetry in the Galaxy or in full 3D.
The boundaries of the model in radius and height, and the grid spacing, are user-definable.
 In addition there is a grid in momentum;
momentum (not e.g. kinetic energy) is used because it is the natural quantity for propagation in equation (1).
Parameters for all processes in equation (1) can be controlled on input.
The distribution of CR sources can be freely chosen, typically to represent SNR.
Source spectral shape and isotopic composition (relative to protons) are input parameters.
Interstellar gas distributions are based on current HI and CO surveys, and the interstellar radiation field is based on
a detailed calculation. 
Cross-sections are based on extensive compilations and parameterizations
\cite{2004AdSpR..34.1288M}. 
The numerical solution proceeds in time until a steady-state is reached; a time-dependent solution is also an option.
Starting with the heaviest  primary nucleus considered (e.g.\ $^{64}$Ni) the propagation solution is used to compute the source term for its spallation products,
which are then propagated in turn, and so on down to protons, secondary electrons and positrons, and antiprotons.
In this way secondaries, tertiaries etc. are included.
(Production of $^{10}$B via the $^{10}$Be-decay channel is  important and requires a second iteration of this procedure.)
 GALPROP includes K-capture and electron stripping
processes, where a nucleus with an electron (H-like) is considered a separate species because of the difference in the lifetime.
Since H-like atoms have only one K-shell electron,
the K-capture decay half-life has to be increased by a factor of 2
compared to the measured half-life value.
 Primary electrons  are treated separately.
Normalization of  protons, helium and electrons to experimental data is provided (all other isotopes  are determined by the source composition and propagation).
\grays and synchrotron are computed  using interstellar gas data (for pion-decay and bremsstrahlung)   and the ISRF model (for inverse Compton).
Spectra of all species on the chosen grid and the \gray and synchrotron skymaps  are output in a standard astronomical format  for comparison with data.
 Recent extensions to GALPROP include non-linear wave damping
\cite{2006ApJ...642..902P} 
  and a dark matter package.

The computing resources required by GALPROP are moderate by today's standards.
We remark that
while GALPROP has the ambitious goal of being `realistic', it is obvious that any such model  can only be a crude approximation to reality.
Some known limitations are:
only  energies below $10^{15}$ eV (no trajectory calculations), uniform source abundances (no superbubble enhancements),
 only scales $>$10 pc (no clumpy ISM: limited by computer power), 
 B-field treated as random for synchrotron (regular component  affects structure of radio emission).
 For these cases other techniques may be more appropriate, and they provide a goal for future developments of GALPROP.


\subsection{Numerical versus analytical} 

The following expresses the authors' opinion on this matter.
The analytical approaches are claimed to have various advantages as follows:

1. {\bf Physical insight:} 
of course   it is true that analytical solutions for simple cases are very useful to get insight into the relations between the quantities
involved, and for rough estimates. In fact the analytical formulae may become so complicated that finally  no insight is gained. In contrast the  numerical models
are very intuitive since they generate explicitly the CR distribution over the Galaxy for all species.
2. {\bf Equivalent to  full solution of propagation equation:} only true under restrictive conditions, especially involving energy-losses and
  spatially varying densities. Electrons and positrons are anyway  beyond analytical methods (for energy-losses on realistic 
interstellar radiation fields), while these are an essential
component of the CR.
3.  {\bf Faster, easier to compute}: with todays computers the speed issue has become irrelevant, and the implementation of a numerical model is not
harder than the complicated integrals over Bessel functions etc.

In summary we can do no better than a quotation from a paper of 26 years ago (!)
\cite{1981ApJ...245..753W}: 
`It is unclear whether one would wish to go much beyond the generalizations discussed above for an analytically soluble diffusion model.
The added insight from any analytic solution over a purely numerical approach is quickly cancelled by the growing complexity of the formulae.
With rapidly developing computational capabilities, one could profitably employ numerical solutions...'
We remark also that for CR air-shower calculations, analytical methods gave way to numerical ones at least 40 years ago.


\subsection{Self-consistent  models}

A few attempts at a self-consistent  description of CR in the Galaxy have been made,
including them as a relativistic gas as one component of ISM dynamics.
This is obviously much harder than the phenomenological models described above
which treat propagation in a prescribed environment.
 3D models of the magnetized ISM with a CR-driven wind have been made by
\citep{1996A&A...311..113Z,1997A&A...321..434P} 
and this is claimed to be also consistent with CR secondary/primary ratios.
Such a wind  has been put forward  as a possible explanation of the CR gradient problem
 \cite{2002A&A...385..216B}. 
The Parker instability has been re-analysed recently
\cite{2000ApJ...543..235H} 
using anisotropic diffusion
\cite{1999ApJ...520..204G} 
and followed by
a  CR-driven Galactic dynamo model
\cite{2004ApJ...605L..33H} 
which uses an extension the Zeus-3D MHD code
\cite{2003A&A...412..331H} 
  including CR propagation and sources.
CR propagation in a magnetic field produced by dynamo action of a turbulent flow
\cite{2006MNRAS.tmp.1221S} 
presents the whole subject from a novel viewpoint.
 The extension of such approaches to include CR spectra, secondaries, \grays etc.,
which would provide a complete set of comparisons with observations,
would be very desirable but has not yet been attempted.
Another kind of self-consistency is to include the effect of CR on
the diffusion coefficient
\cite{2006ApJ...642..902P} 
as described in more detail in Section 2.5.


\section{Confrontation of Theory with Data}

\subsection{Stable secondary/primary ratios}

The reference ratio is almost always B/C because B is entirely secondary,  
the measurements are better than for other ratios and are available up to 100 GeV.
 Because C,N,O are the major progenitors of B, the production cross sections are better known
than, e.g. in the case of Be and Li 
\cite{2003ICRC....4.1969M,2006AdSpR..38.1558D}.  

 The usual procedure is to use a leaky-box or weighted-slab  formalism with the empirical rigidity-dependence
 $X(R)=(\beta/\beta_0) X_0$, $(\beta/\beta_0)(R/R_0)^{-\alpha} X_0$
for $R<R_0$, $R>R_0$ respectively.
The break at $R_0$ is required because B/C is observed to decrease to low energies faster than the $\beta$-dependence
(which just describes the velocity effect on the  reaction rate).
The source composition depends on the form and  parameters of  $X(R)$, and vice-versa,
(since for example B is produced by C,N,O etc)
 so the procedure is iterative, starting from a solar-like composition.

\begin{figure}
\centerline{\psfig{figure=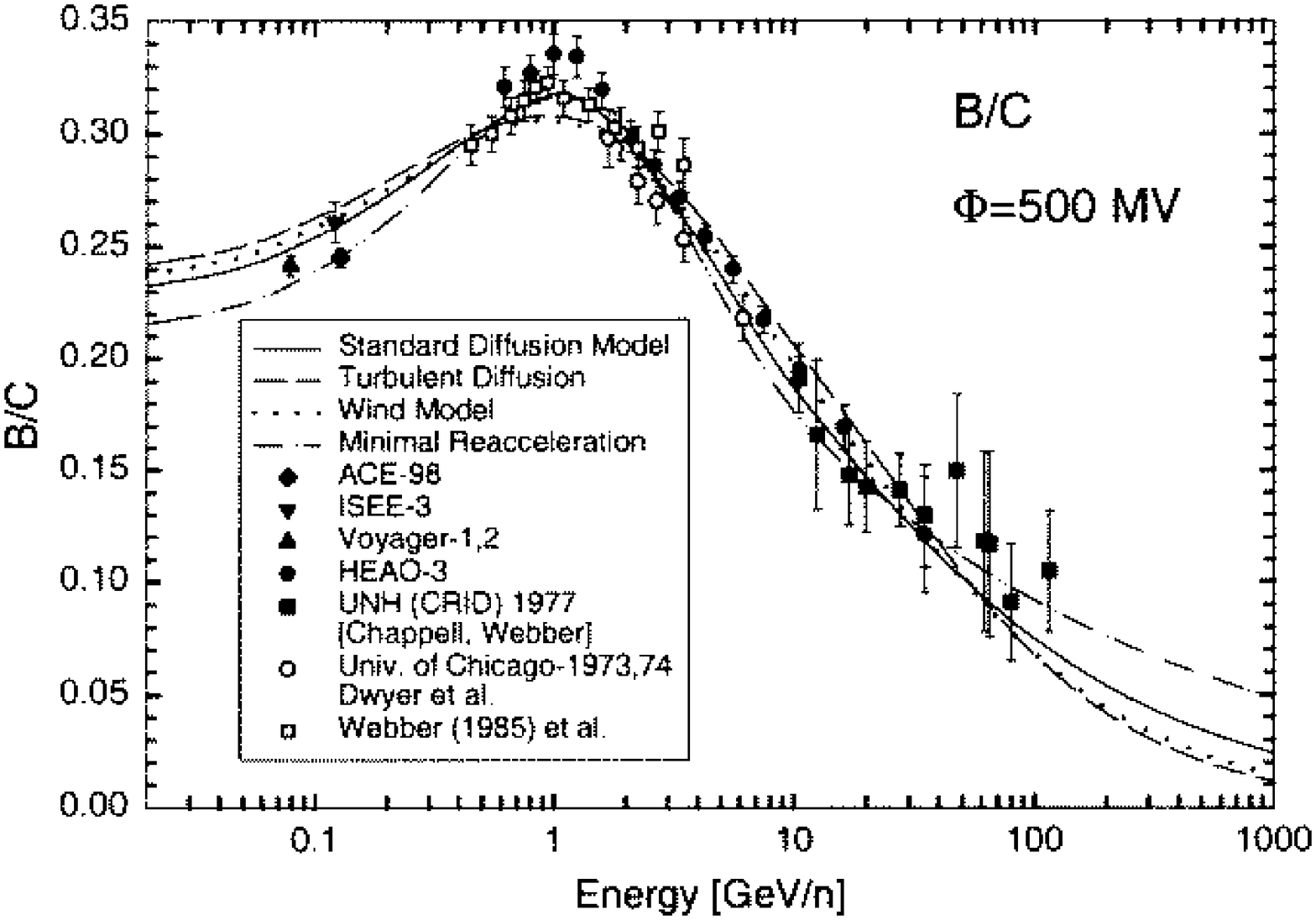,height=14pc}
            \psfig{figure=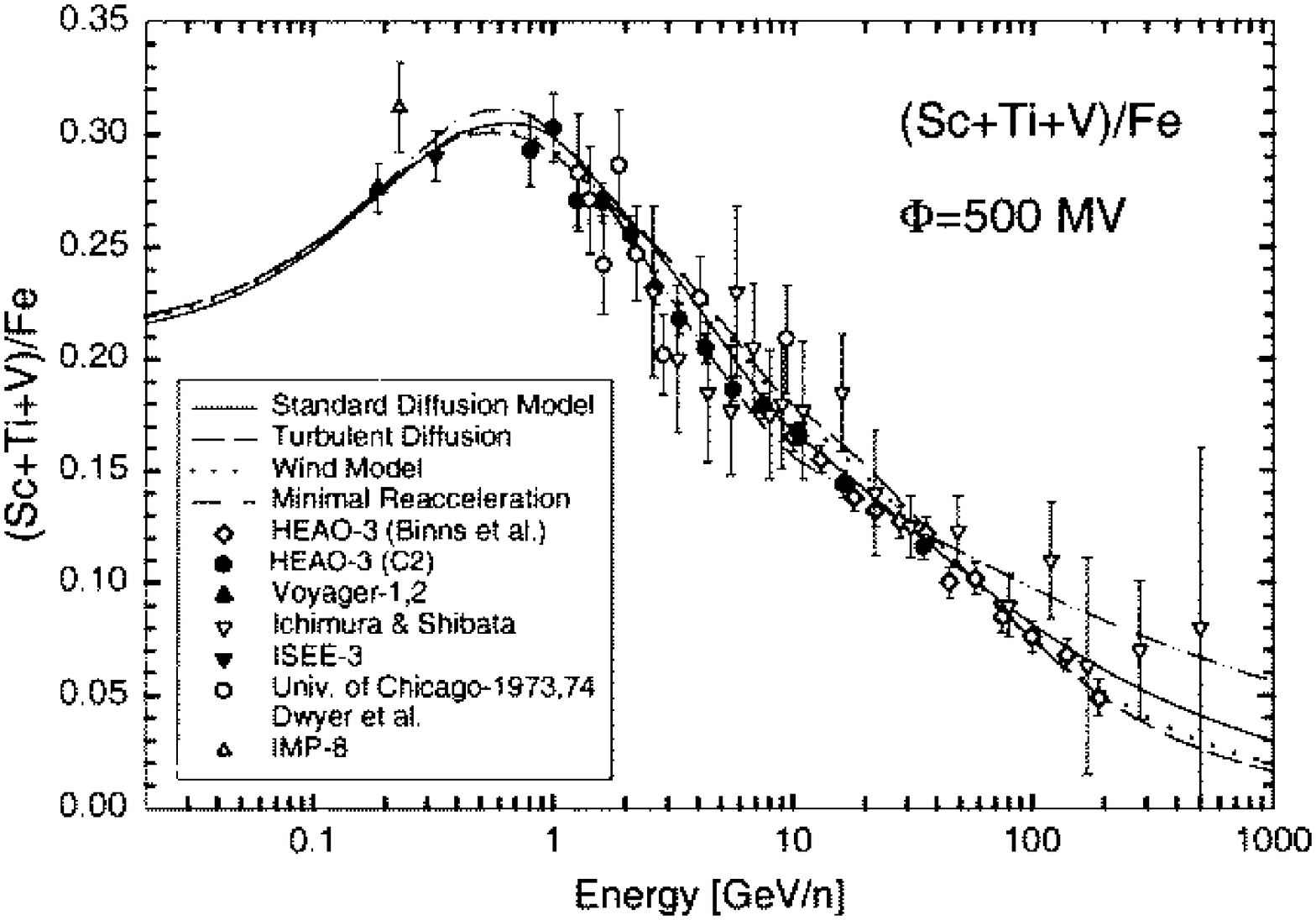,height=14pc}}
\caption{B/C and sub-Fe/Fe data compilation compared to  four models treated by the modified weighted-slab technique, from  
\citep{2001ApJ...547..264J}. 
 }
\label{BC_subFe_Jones}
\end{figure}

\begin{figure}
\centerline{\psfig{figure=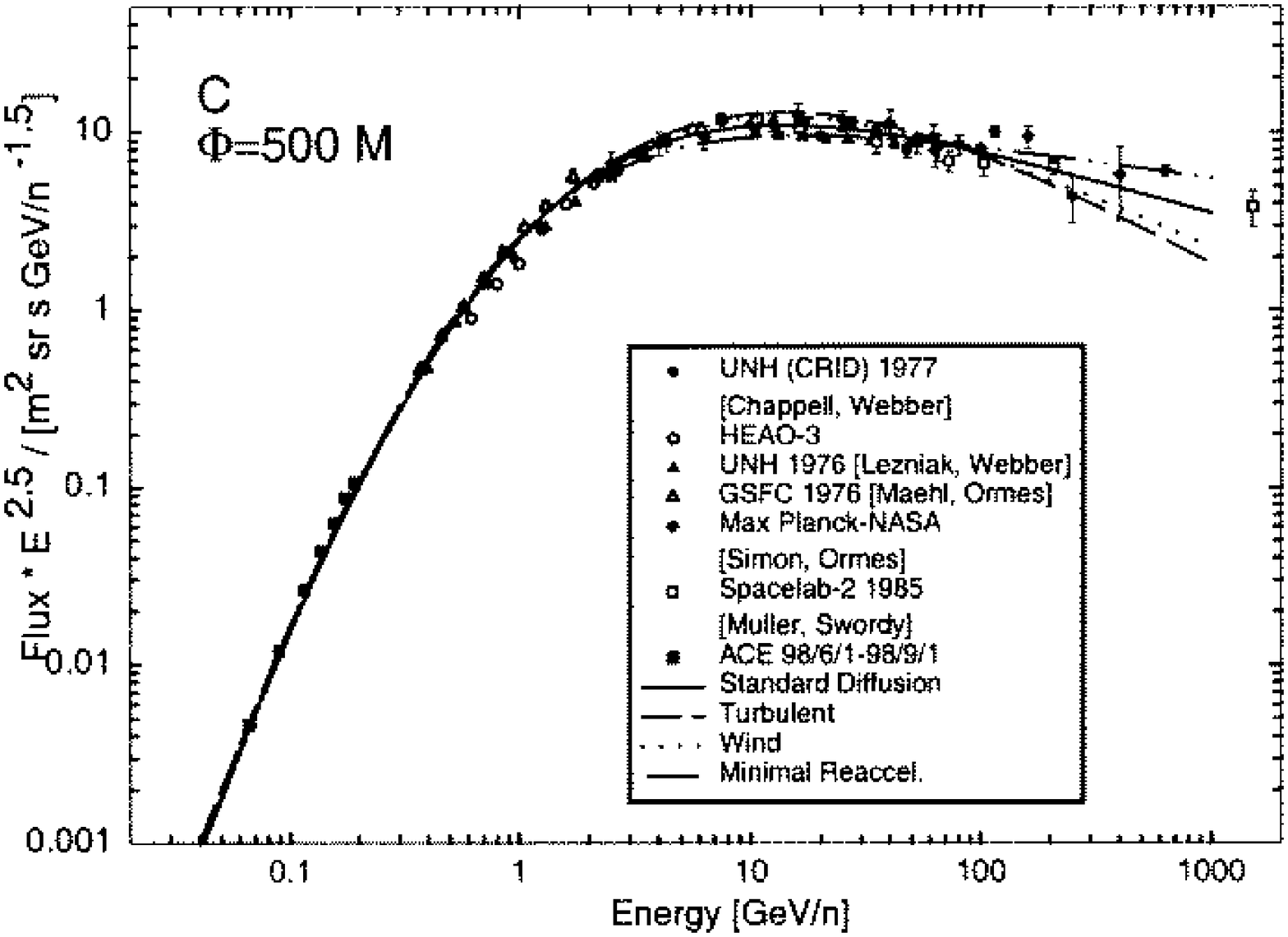,height=14pc}
            \psfig{figure=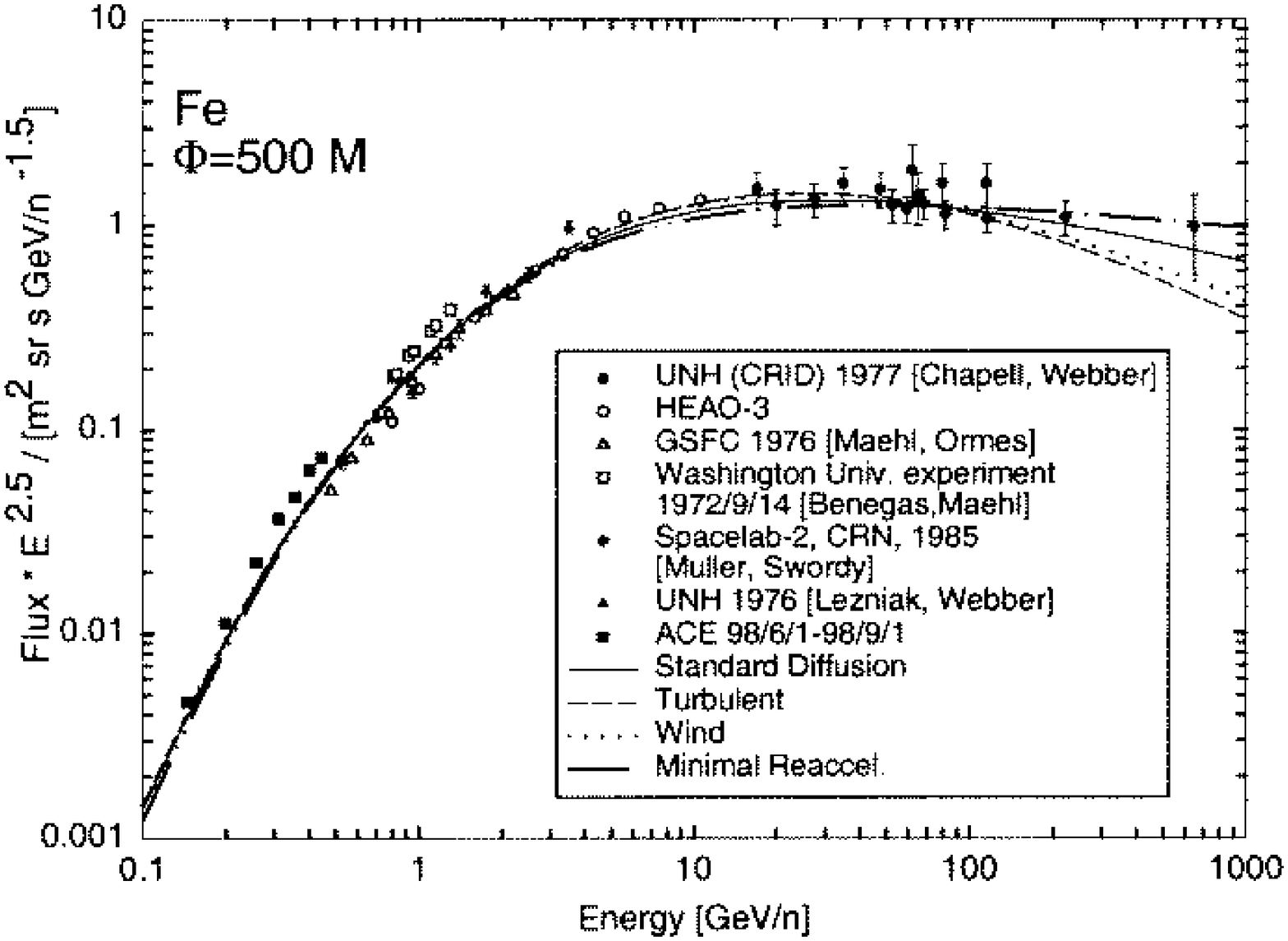,height=14pc}}
\caption{Data compilation for spectra of C and Fe compared to  four models treated by the modified weighted-slab technique, from  
\citep{2001ApJ...547..264J}. 
 }
\label{C_Fe_Jones}
\end{figure}

 A typical parameter set fitting the data 
\citep{2001ApJ...547..264J} 
is  $\alpha$ =  0.54 , $X_0$ = 11.8 g cm$^{-2}$, $R_0$ = 4.9 GV/c, with a source spectrum rigidity index --2.35.
In principle all other secondary/primary ratios should be consistent with the same parameter set.
This is generally found to be  the case.
As a state-of-the-art application of the  weighted slab technique of
\citep{1996ApJ...465..972P} 
 we again refer to
\citep{2001ApJ...547..264J}. 
This is applicable to stable nuclei only but includes energy losses and gains subject to the limitations described in Section 2.8.
They apply the method to 1-D disk-halo diffusion, convection, turbulent diffusion and reacceleration models cast in weighted-slab form.
Fig.~\ref{BC_subFe_Jones} shows B/C and 
(Sc+Ti+V)/Fe, so-called sub-Fe/Fe, 
from their paper. Clearly the models cannot be distinguished based on these types of data alone,
and they can all provide an adequate fit; this shows the importance of using other species as well
as the ones used here.
 The models can  be used to obtain the injection spectrum of primaries,
and they find an index of 2.3 to 2.4 for C and Fe in the energy range 0.5 -- 100 TeV, with the propagated spectrum and data shown in
Fig.~\ref{C_Fe_Jones}.

It has been claimed that no break in  $X(R)$ is required to fit Voyager 2 outer-heliosphere
 B/C, N/O and sub-Fe/Fe  data extended to 1.5 GeV, plus HEAO3 data, and adopting suitable solar modulation levels
\citep{2003ApJ...599..582W}. 
Voyager 2 provides  a unique dataset because of the low solar modulation. 

We consider now  {\it explicit} models in the sense of Section 2.9.
Basically the same procedure is adopted, with $D_{xx}(R)$ replacing $X(R)$.
Again an ad-hoc  break in  $D_{xx}(R)$ is required in the absence of other mechanisms.
Because of the unphysical nature of such a  $D_{xx}(R)$, many attempts to find a better explanation have been made, including
convection, reacceleration/wave damping,  and local sources.

Convection implies an energy-independent escape from the Galaxy so that it dominates at low energies
as the diffusion rate decreases, so it  gives a simple low-energy asymptotic    $X(R)\propto\beta$.
However  this does not resemble the observed B/C energy-dependence, being too monotonic 
\cite{1998ApJ...509..212S}. 
Furthermore quite severe limits on the convection velocity come from unstable nuclei.

Reacceleration affects the secondary-primary ratio as described in Section 2.5.
Many papers have shown how B/C and other ratios can be reproduced with reacceleration at a
plausible level and no ad-hoc break in the diffusion coefficient. An example is shown in Fig.~\ref{BC_subFe_Jones}.
An application  including recent ACE (Li, Be, B, C) data is in
\cite{2006AdSpR..38.1558D}. 
Since reacceleration at some level {\it must} be present if diffusion occurs on moving scatterers (e.g. Alfv\'en waves)
this mechanism is favoured but it is not proven. Direct evidence for reacceleration could come
from certain K-capture nuclei (Section 3.4).                   
Reacceleration requires a smaller value of  $\alpha$,  typically 0.3 -- 0.4 ,  consistent with Kolmogorov turbulence, which
helps solve the problems with anisotropy (section 3.5).

\begin{figure}
\centerline{
\psfig{figure=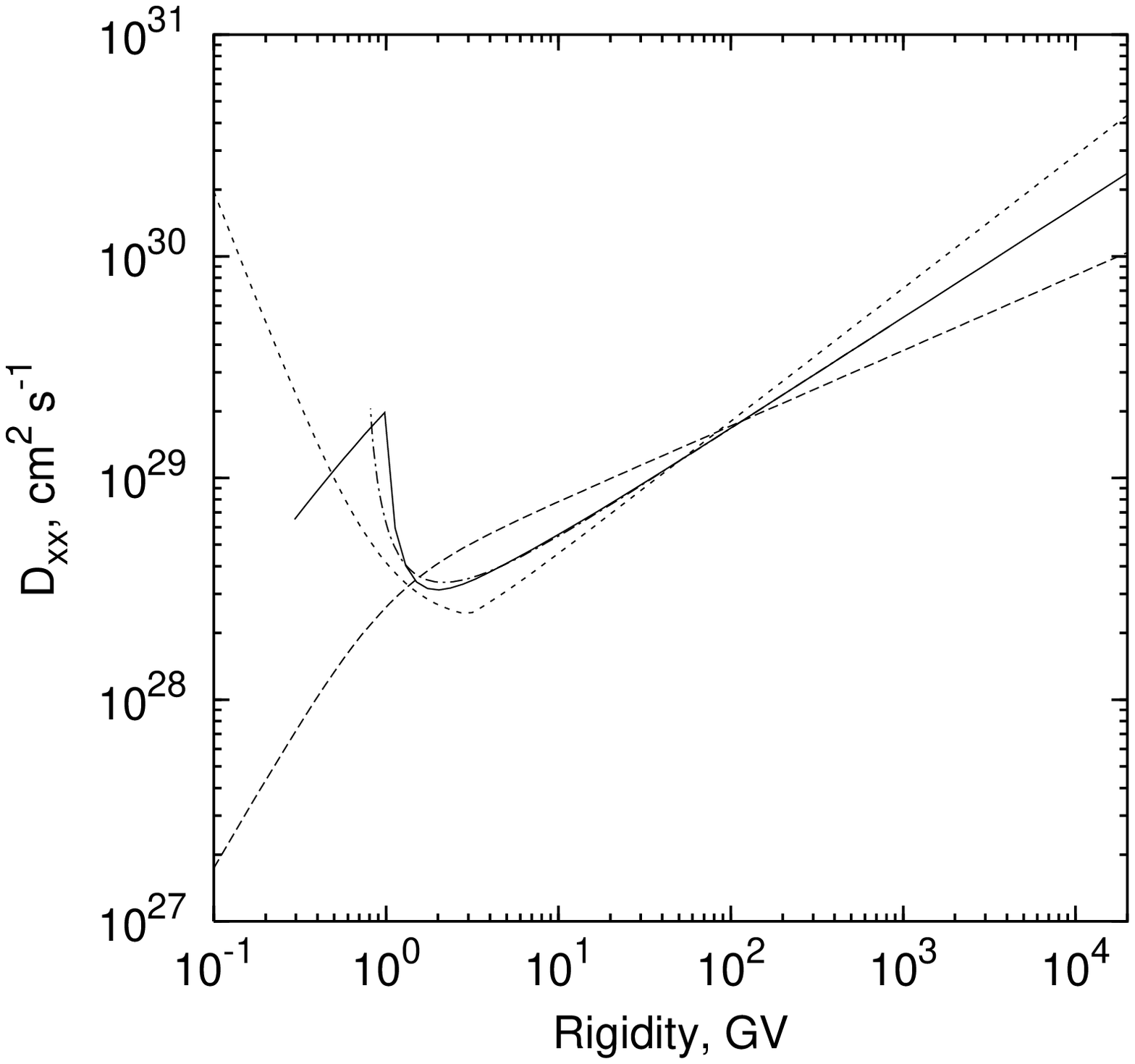,height=11pc}
\psfig{figure=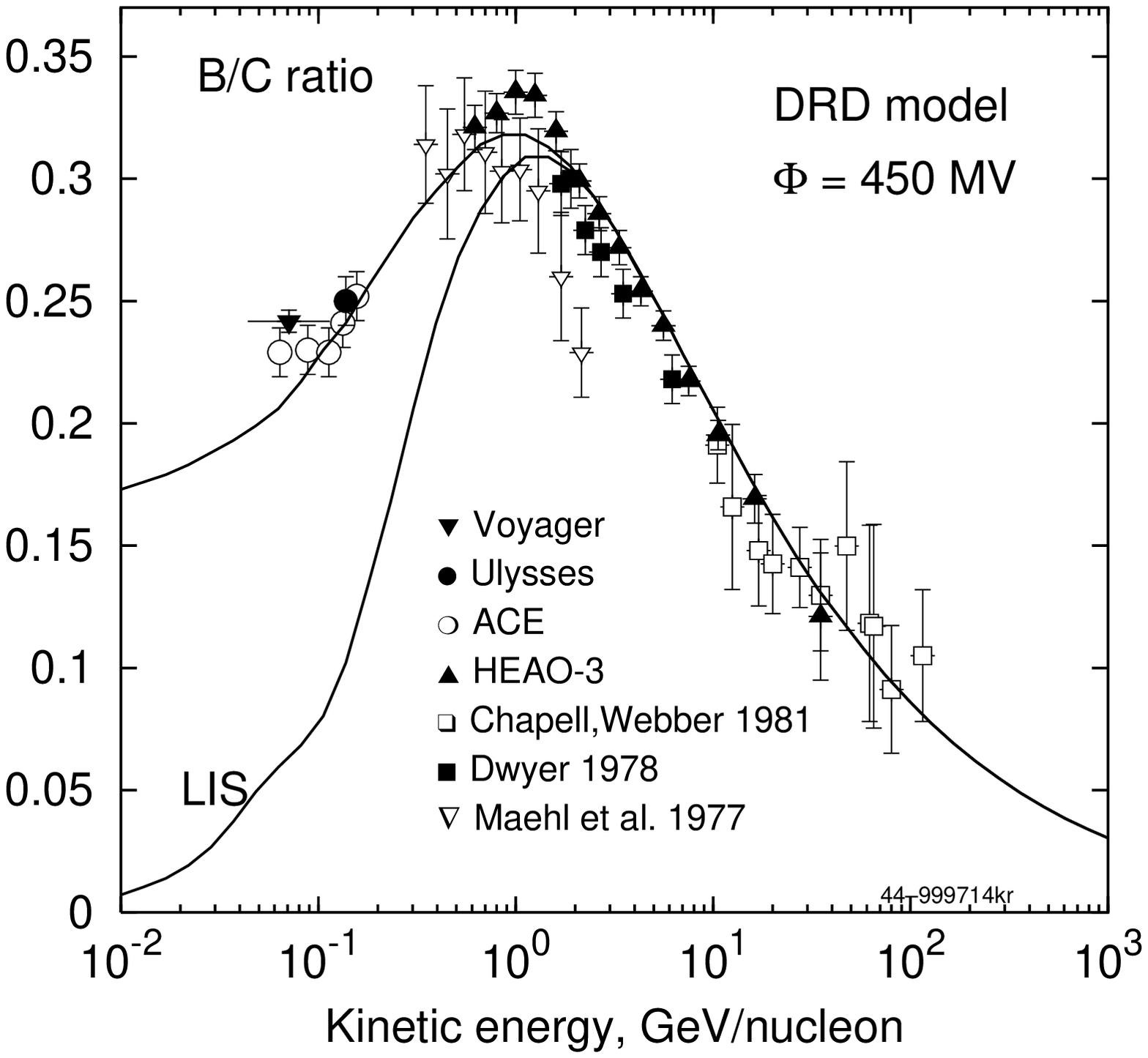,height=11pc}}
\centerline{
\psfig{figure=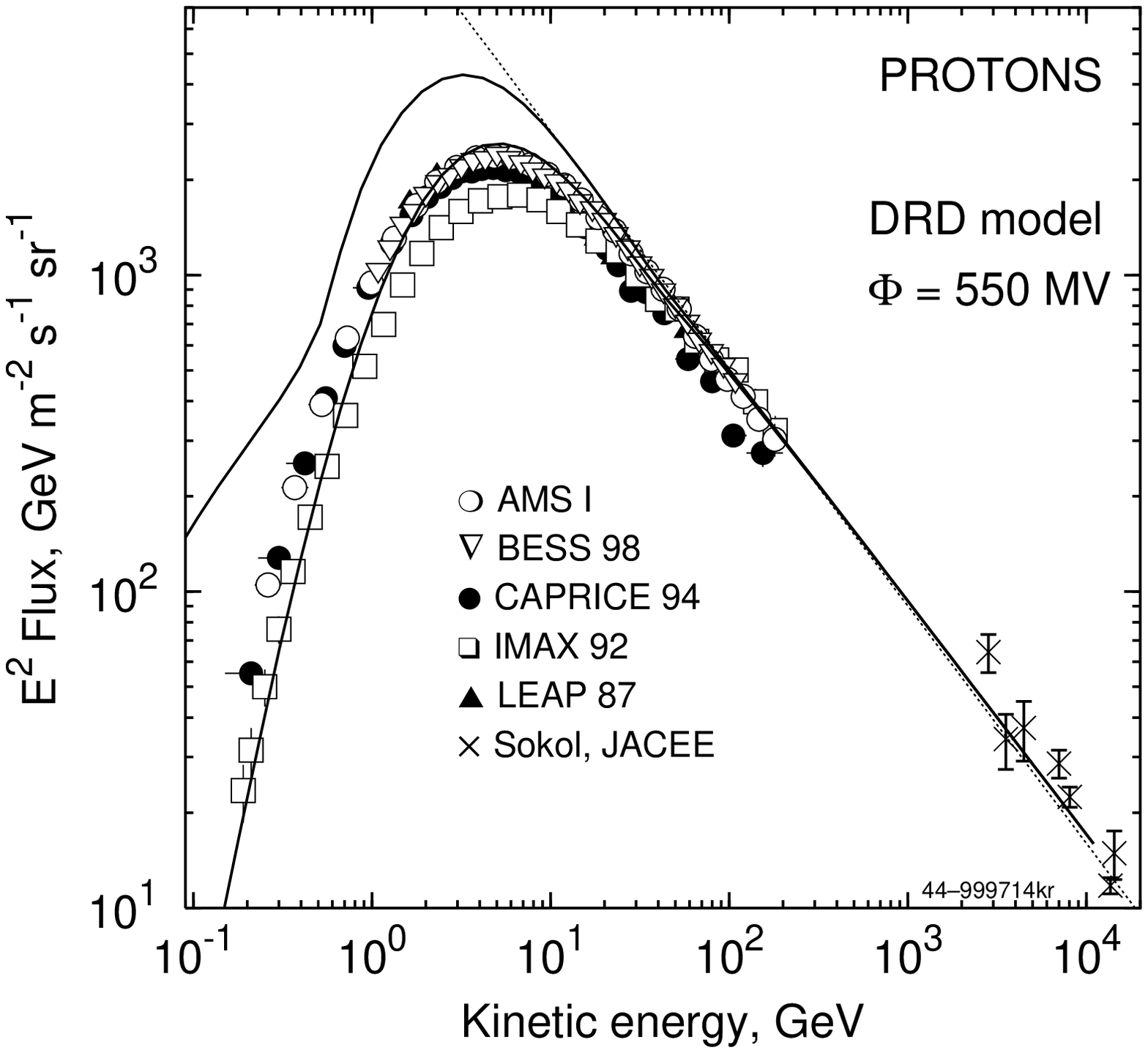,height=11pc}
\psfig{figure=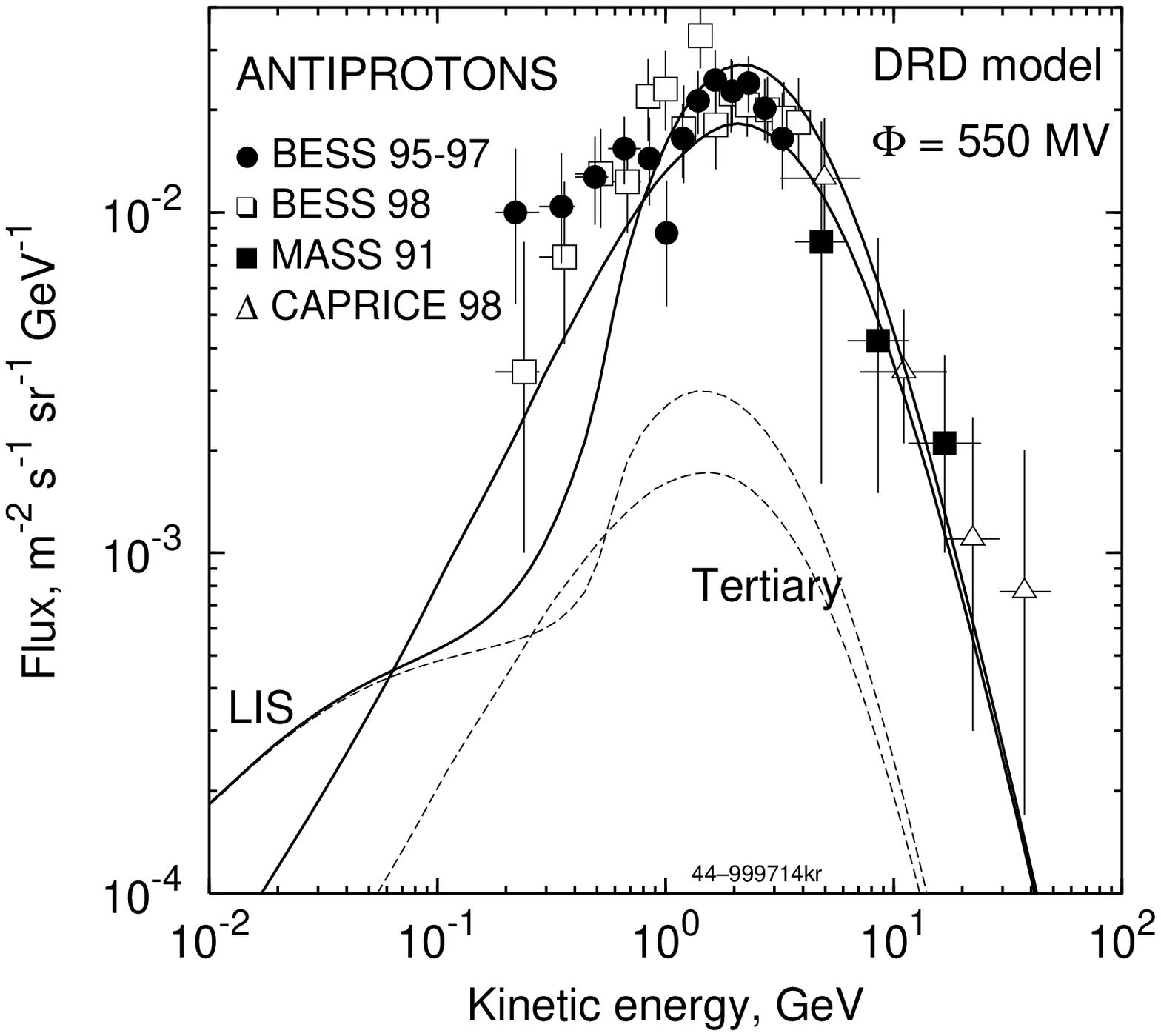,height=11pc}}
 \caption{
Upper left:
diffusion coefficient in different models 
\cite{2006ApJ...642..902P}: 
plain diffusion (dots), Kolmogorov reacceleration (dashes), and Kraichnan-type reacceleration with wave-damping (solid).
Upper right: B/C, lower left: protons, lower right:  antiprotons,  in the wave-damping model 
\cite{2006ApJ...642..902P}. 
LIS marks the local interstellar spectrum. 
 }
\label{BC_p_pbar_damping_Ptuskin}
\end{figure}

Closely related to reacceleration is wave-damping 
as described in Section 2.5.
This can reproduce B/C, protons and antiprotons satisfactorily as shown in Fig.~\ref{BC_p_pbar_damping_Ptuskin}, and also other data 
\cite{2006ApJ...642..902P}. 
 The result of this process is a very sharp rise of the diffusion coefficient at rigidities
 less than about $1.5$ GV.
The Kolmogorov-type dependence is not very successful in this scheme, while a Kraichnan-type works better
 with a  high rigidity asymptotic $D\sim R^{0.5}$ (first panel in Fig.~\ref{BC_p_pbar_damping_Ptuskin}).

A quite different set of parameters has been proposed
\citep{2002A&A...394.1039M}: 
$\alpha=0.7-0.9$ and injection index $\approx2.0$, based on fitting many species simultaneously,
and is suggested to produce the  B/C low-energy decrease from convection. 
Such a large $\alpha$ would give problems for the anisotropy (Section 3.5).
A related analysis
\citep{2001ApJ...555..585M} 
 claims to exclude the Kolmogorov-type spectrum.

An alternative explanation for  the falloff in B/C at low energies invokes weakly nonlinear
(in contrast  to quasi-linear) transport theory of cosmic rays in  turbulent Galactic magnetic fields
\citep{2004ApJ...616..617S,2005ApJ...626L..97S}. 

Another, simpler, explanation of the B/C energy-dependence is the local-source model
\citep{2000AIPC..528..421D,2003ApJ...586.1050M} 
in which  part of the primary CR have an additional local component.
Since the secondary flux has to come from the Galaxy-at-large (the local secondaries being negligible)
a steep local primary source will cause B/C to decrease at low energies.
The known existence of the local bubble containing the Sun, and its probable origin in a few supernovae
in the last few million years, make this a plausible possibility, but hard to prove or disprove.
It is  claimed
\citep{2000AIPC..528..421D} 
 that if B/C is fitted in such a model, then sub-Fe/Fe is not fitted, but an acceptable fit to this and other data {\it is} found in
\cite{2003ApJ...586.1050M}  
using a diffusion model for the large-scale component.


\subsection{Unstable secondary/primary ratios: `radioactive clocks'} 

\begin{figure}%
\centerline{
\psfig{figure=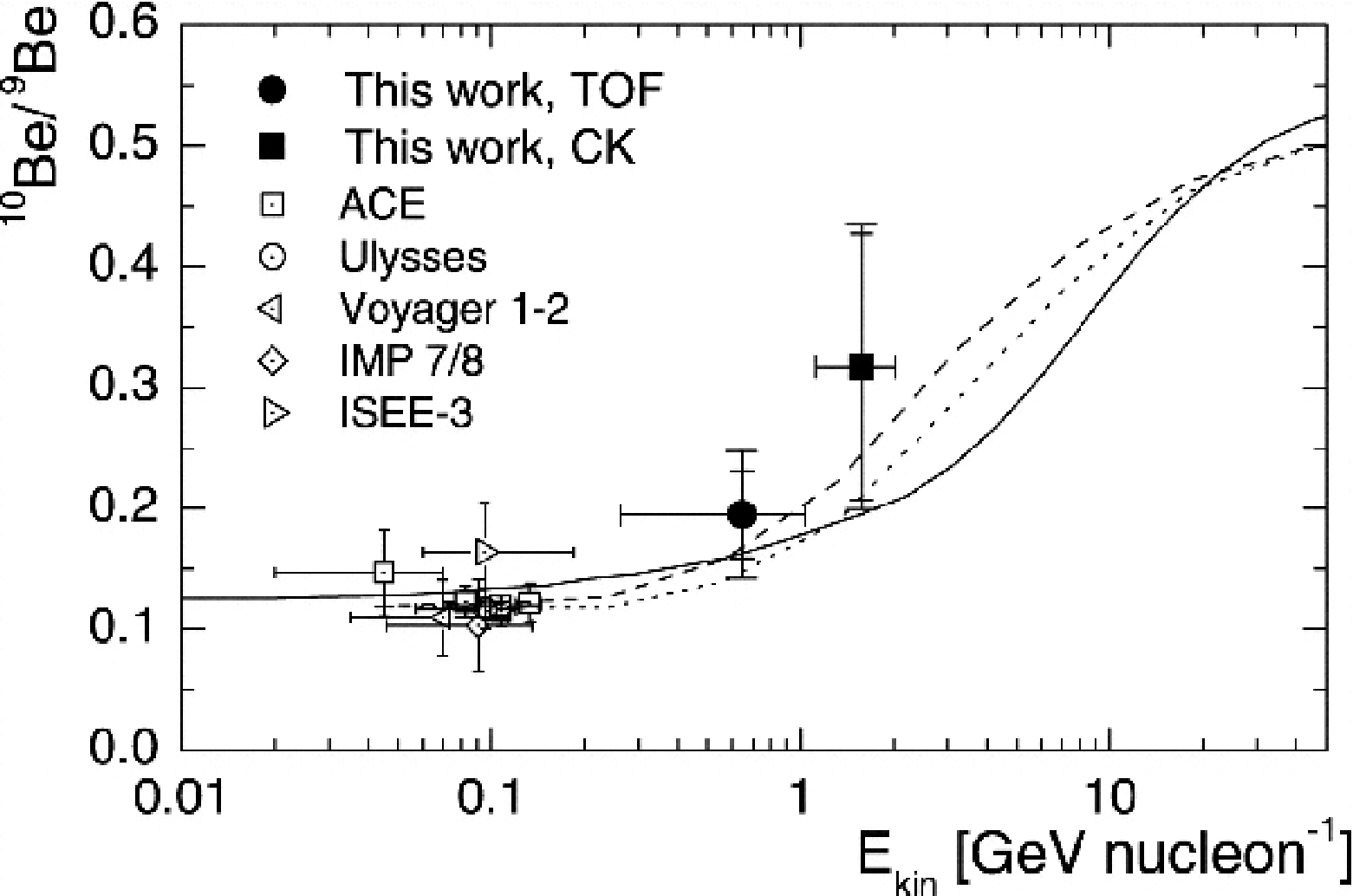,width=9cm}
}
\caption{ Data on energy-dependence of $^{10}$Be/$^{9}$Be including ACE, Ulysses, Voyager, IMP, ISEE-3 and ISOMAX data,
from 
\citep{2004ApJ...611..892H}. 
The solid line is a diffusive halo model with 4 kpc scale height using GALPROP
\citep{2001AdSpR..27..717S} 
 the other lines are leaky-box models
\cite{2001AdSpR..27..743S}. 
}
\label{10Be_Hams}
\end{figure}

The five unstable secondary nuclei which live long enough to be useful probes of cosmic-ray propagation are $^{14}$C, $^{10}$Be, $^{26}$Al, $^{36}$Cl and  $^{54}$Mn,
with properties summarized in 
\cite{1987ApJS...64..269G,2002A&A...381..539D,2003ApJ...586.1050M}. 
 $^{10}$Be is the longest-lived and best measured. The theory is presented in section 2.2.
Based on these isotopes and updated cross-sections
\citep{2001ICRC....5.1836M} 
find $z_h$=4--6 kpc, consistent with their earlier estimates of 3--7 kpc
\citep{2001AdSpR..27..717S} 
and 4--12 kpc
\citep{1998ApJ...509..212S}. 
Fig.~\ref{10Be_Hams} shows a comparison of  $^{10}$Be/$^{9}$Be  with models; the ISOMAX  $^{10}$Be measurements
\citep{2004ApJ...611..892H} 
 up to 2 GeV (and hence longer decay lifetime) are consistent
with the fit to the other data, although the statistics are not very constraining.

The data are often interpreted in terms of the leaky-box model, but this is misleading 
\citep{1998A&A...337..859P,1998SSRv...86..225P,2002A&A...381..539D}. 
For the  formulae and the detailed procedure for the  leaky-box model interpretation see
\cite{2001ApJ...563..768Y}. 
Luckily the  leaky-box model surviving fraction can be converted to a physically meaningful quantities 
\cite{1998SSRv...86..225P} 
 for a given model;
for example in a simple diffusive halo model, the surviving fraction determines the diffusion coefficient,
 which can be combined with stable secondary/primary ratios to derive the halo size.
Typical results are $D_{xx} = (3-5)\times 10^{28}$ cm$^2$ s$^{-1}$ (at 3 GV) and $z_h = 4$ kpc.
We can then compare the  leaky-box model  `escape time' $\approx10^7$ yr with the actual time for CR to reach the halo boundary after leaving their sources, the latter being typically an order of magnitude larger.
The  leaky-box model `gas density' is typically 0.3 cm$^{-3}$ compared to the actual average density 0.03 cm$^{-3}$  for a 4 kpc halo height, again an order-of-magnitude difference.

Since radioactive secondaries only travel a few hundred pc before decaying it is sometimes considered
\citep{2001SSRv...99...27M} 
 that they cannot
give information on the propagation region of CR; this is somewhat misleading since it is precisely the {\it combination} of stable and radioactive
data which {\it does} allow this; the radioactives determine the diffusion coefficient, which then allows the
 size of the full propagation region to be determined from the stable 
secondary/primary ratio (where the CR do come from the entire containment region). This of course assumes the  diffusion coefficient
does not vary greatly from the local region to the full volume.

The effect of the local bubble  surrounding the Sun  on the interpretation of radioactive nuclei was pointed out by 
\citep{2002A&A...381..539D}. 
The flux of unstable secondaries is reduced if they are not produced in a gas-depleted region around the Sun,
since they will decay before reaching us,
and this could lead to an overestimate of the halo size if interpreted in a simple diffusive halo model.
 This effect  would be reduced if  the diffusion coefficient in the local region were larger than the large-scale value.
In fact the situation is even more complex.
According to
\citep{2006astro.ph..6743F,2006sjsg.book.....F,2006astro.ph..7600M} 
the Sun left the local bubble about $10^5$ years ago after spending several million years inside,
and we now live in the CLIC (collection of local interstellar clouds) with HI density about 0.2 cm$^{-3}$ and 35 pc extent.
This aspect of the problem for CR propagation  has not yet been addressed.

\subsection{K-capture isotopes and acceleration delay} 

Three  isotopes produced in explosive nucleosynthesis decay essentially only by
K-capture: $^{59}$Ni ($7.6\times 10^4$ y), $^{57}$Co (0.74 yr), $^{56}$Ni (6d). If acceleration occurs
before decay, the decay will be suppressed since the nuclei are stripped.
$^{56}$Ni is absent  as expected, but the other two nuclei are more interesting.
\citep{2000AIPC..528..363W} 
used ACE data on these nuclei to show that the delay between synthesis and acceleration
is long compared with the $^{59}$Ni decay-time, unless significant $^{59}$Co is synthesised in 
supernovae; considering theoretical  $^{59}$Co yields they conclude on a delay $\ge 10^5$ y.
This is inconsistent with models in which supernova accelerate their own ejecta,
but consistent with acceleration of existing interstellar material.
The possibility of in-flight electron attachment complicates the analysis however 
\citep{2000AIPC..528..363W}. 
For more  discussion see
 \citep{2001SSRv...99...27M}. 
A result from TIGER on Co/Ni at 1-5 GeV/n 
\cite{2005ICRC....3...61D} 
 supports the acceleration delay also at higher energies.
\subsection{K-capture isotopes and reacceleration} 

Analyses using ACE data for $^{49}$V, $^{51}$V, $^{51}$Cr, $^{52}$Cr and  $^{49}$Ti  and other nuclei    have been made  
\citep{2000AIPC..528..406N}, 
\citep{2001ICRC....5.1675N}, 
\citep{2001AdSpR..27..737J}. 
\citep{2001ICRC....5.1675N} 
found that while $^{51}$V/$^{52}$Cr was in better agreement with reacceleration models, $^{49}$Ti/$^{46+47+48}$Ti gave
the opposite result.
\citep{2001AdSpR..27..737J} 
find that while V/Cr ratios are in slightly better agreement with models including reacceleration,
the  ratios involving Ti are inconclusive.
\citep{2001ICRC....5.1675N} 
used ACE data on $^{37}$Ar, $^{44}$Ti, $^{49}$V, $^{51}$Cr, $^{55}$Fe and $^{57}$Co,  with inconclusive results.
The main problem is the accuracy of the fragmentation cross-sections
 \cite{2003ApJ...586.1050M}.  
Discussion of ACE and previous results can be found in section 3 of 
 \citep{2001SSRv...99...27M}. 


\subsection{Anisotropy}

\begin{figure}
\centerline{
\psfig{figure=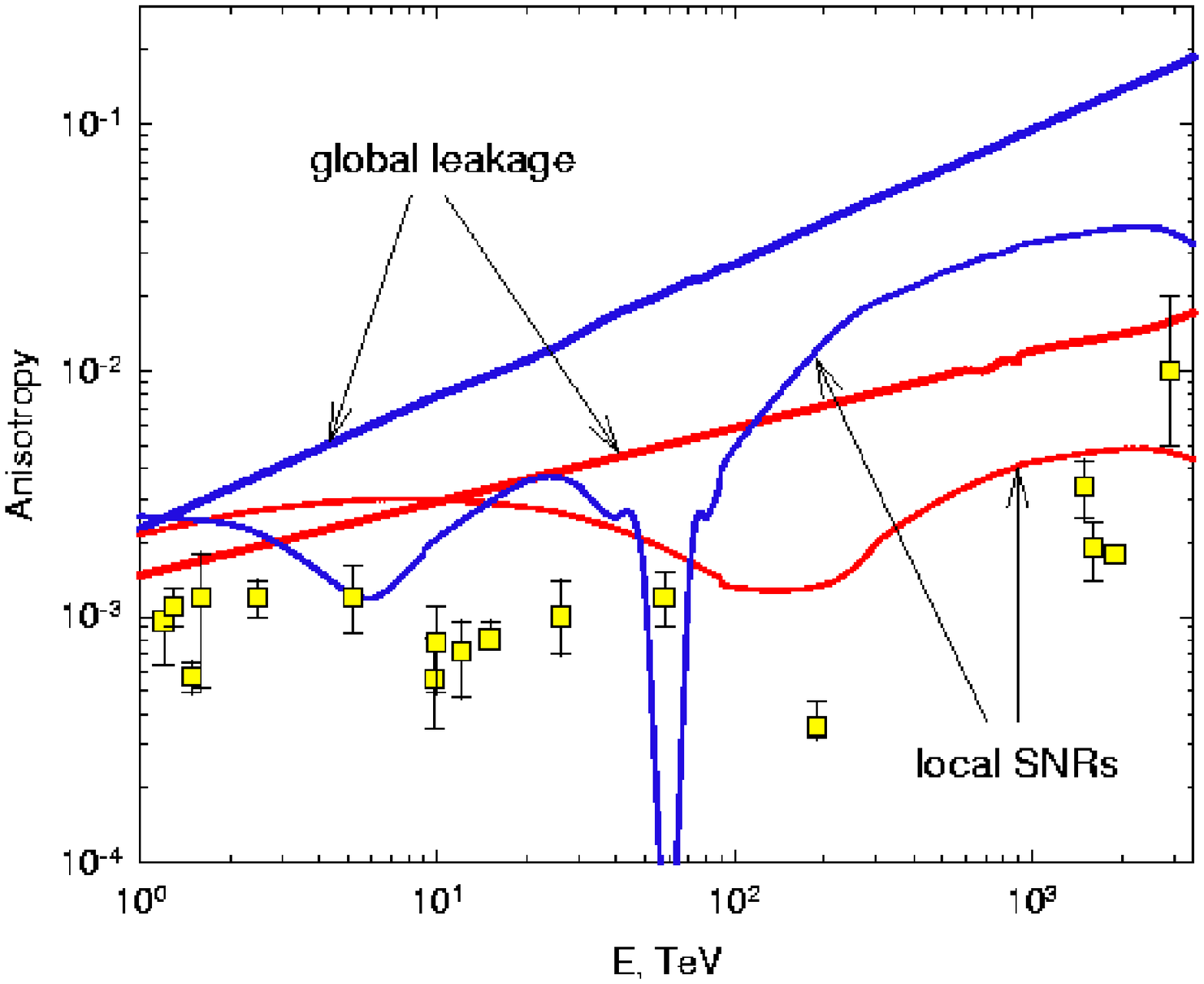,height=20pc}
}
\caption{The anisotropy of cosmic rays in the reacceleration (red curves)
and the plain diffusion (blue curves) models. Separately shown are the effects of
the global leakage from the Galaxy (thick lines), and the contribution from
local SNR (thin lines). The collection of data on cosmic ray anisotropy
(yellow squares) are taken from
 \cite{2003PhRvD..67d2002A} 
 where the references to
individual experiments can be found.}
\label{AnisotropyFig}
\end{figure}

High isotropy is a distinctive quality of Galactic CR observed at
the Earth. The global leakage of CR from the Galaxy and the contribution
of individual sources lead to anisotropy but the trajectories of energetic
charged particles are highly tangled by regular and stochastic
interstellar magnetic fields which isotropize the CR angular
distribution. This  makes difficult or even impossible the direct association
of detected CR  particles with their sources,  except for the highest
energy particles.  Observations give  the amplitude of the first
angular harmonic of anisotropy at the level of $\delta \sim 10^{-3}$ in the
energy range $10^{12}$ to $10^{14}$~eV where the most reliable data are
available, see
 \cite{2005ApJ...626L..29A,2006NuPhS.151..485T} 
 and Fig.~\ref{AnisotropyFig}. The angular
distribution of particles at lower energies is significantly modulated by
the solar wind. The statistics at higher energies are not good enough yet but
the measurements indicate the anisotropy amplitude at a level of a few
percent at $10^{16}-10^{18}$~eV.  The data of the Super-Kamiokande-I detector
\cite{2005ICRC....6...85G}  
 allowed   accurate two-dimensional mapping of 
CR  anisotropy at $10^{13}$ eV (see also the Tibet Air Shower Array results
 \cite{2005ApJ...626L..29A}). 
 After correction for atmospheric effects, the  deviation
from the isotropic event rate is $0.1\%$ with a statistical significance of
$>$$5\sigma $ and  direction to maximum excess at roughly $\alpha =75^\circ$,
$\delta =-5^\circ$.

The amplitude of anisotropy in the diffusion approximation is given by the
following equation which includes  pure diffusion and  convection terms: 
$\delta=-[3D\mathbf\nabla f+\mathbf{u}p\,(\partial f/\partial p)]/vf$, 
see \citep{1990acr..book.....B}. 
%
Here $D$ is the diffusion tensor, and it is assumed that the magnetic
inhomogeneities which scatter CR  particles are frozen in the
background medium, moving with  velocity $u\ll v$, which gives rise to the
convection term (also called the Compton-Getting term).

The Compton-Getting anisotropy is equal to $(\gamma +2)u/c$ for
ultra-rerelativistic CR with a power law spectrum $I(E)\sim
p^{2}f(p)\sim E^{-\gamma }$. The motion of the Solar System through the
local interstellar medium produces the constant term in the energy-dependence of the  anisotropy $\sim$$4\times 10^{-4}$, with a maximum intensity in the general direction to the
Galactic centre region, which  does not agree with the data at $10^{12}-10^{14}$
eV which point to an  excess intensity from the anticenter hemisphere. The
convection effect is outweighed by the diffusion anisotropy which is due to
the non-uniform distribution of CR in the Galaxy. The systematic
decrease of CR intensity to the periphery of the Galaxy and the
CR fluctuations produced by  nearby SNR  are approximately
equally important for the formation of the local CR gradient.

Calculations of CR anisotropy
\cite{2006AdSpR..37.1909P} 
are illustrated in
Fig.~\ref{AnisotropyFig} where the effect of global leakage
from the Galaxy and the overall contribution of known SNR with distances up
to $1$ kpc are shown separately.
Two basic versions of the flat-halo diffusion model -- the plain diffusion
model with $D\sim E^{0.54}$ and the model with reacceleration where $\Dxx\sim
E^{0.3}$ -- were used in the calculations. The values of $\Dxx$ were taken from %
\citep{2001ApJ...547..264J}. 
The diffusion from a few nearby SNR with different ages and distances
results in a non-monotonic dependence of anisotropy on energy. The flux
from the Vela SNR probably dominates over other SNR at energies below about 
$6\times 10^{13}$ eV in the plain diffusion model and below about $10^{14}$
eV for the model with reacceleration. The results of these calculations
indicate that the diffusion model with reacceleration is compatible with the
data on CR anisotropy  within a factor of about $3$. On the other
hand it seems that the plain diffusion model with its relatively strong
dependence of diffusion on energy predicts a too large anisotropy at $%
E>10^{14} $ eV. It should be stressed that CR diffusion is here
assumed to be isotropic, which is a seriously simplifying assumption but which is
difficult to avoid because of the complicated and in fact unknown  detailed
structure of the Galactic magnetic field. The presence of a large-scale random
magnetic field justifies the approximation of isotropic diffusion on
scales larger than a few hundred parsecs, but the anisotropy is a
characteristic which is very sensitive to the local surroundings of the Solar
System including the direction of magnetic field and the value of diffusion
tensor, see discussion in \citep{1990acr..book.....B}. 


\subsection{Diffuse Galactic Gamma Rays}

Gamma rays (above about 100 MeV)  from the interstellar medium hold great promise for CR studies since they originate
throughout the Galaxy, not just the local region of direct measurements.  The complementarity
of  \grays and direct measurements  can be exploited to learn most about CR origin and propagation.
Despite this the interpretation has brought surprises in that the \gray spectrum is not just as
would be expected from the directly-observed CR spectra.
 \grays are produced in the interstellar medium by interactions of CR protons and He ($\pi^0$-decay) and electrons (bremsstrahlung) with gas, 
and electrons with the interstellar radiation field via inverse-Compton scattering.
For details of the processes the reader is referred to 
\citep{1998ApJ...493..694M,2000ApJ...537..763S,2000ApJ...528..357M,2004ApJ...613..962S}. 
Additional astronomical material comes into play like the distribution of atomic and molecular gas, and the interstellar radiation field.
In fact  \grays provide an important independent handle on molecular hydrogen  and its relation to its CO molecular tracer, which has taken its place
beside more traditional determinations.

Historically  observations started with the OSO-III satellite in 1968, followed by SAS-2 in 1972, COS-B  (1975--1982) and CGRO (1991--2000).
Each of these experiments represented a significant leap forward with respect to its predecessor. 
SAS-2 established the existence of emission from the interstellar medium and allowed a first attempt to derive the CR distribution.
With COS-B the CR distribution could be better derived and it was found  not to follow the canonical distribution of
supernova remnants, which posed a problem for the SNR origin of CR. A rather dependable value for the CO-H$_2$ relation was derived.
With EGRET and COMPTEL on the Compton Gamma Ray Observatory (CGRO) the improvement in data quality was sufficient to allow such studies to be performed in much greater detail.
There is now so much relevant information and theory that a rather `realistic'  approach is justified and indeed necessary.
At this point the best approach seems to be explicit modelling of the high-energy Galaxy putting in concepts from
CR sources and propagation,  Galactic structure, etc. The idea of a single model to reproduce
both CR and gamma-ray (and other) data simultaneously arose naturally and is the goal of the  GALPROP  project (see Sec 2.10).
For a recent review see \cite{2005AIPC..801...57M}. 

\begin{figure}
\centerline{\psfig{figure=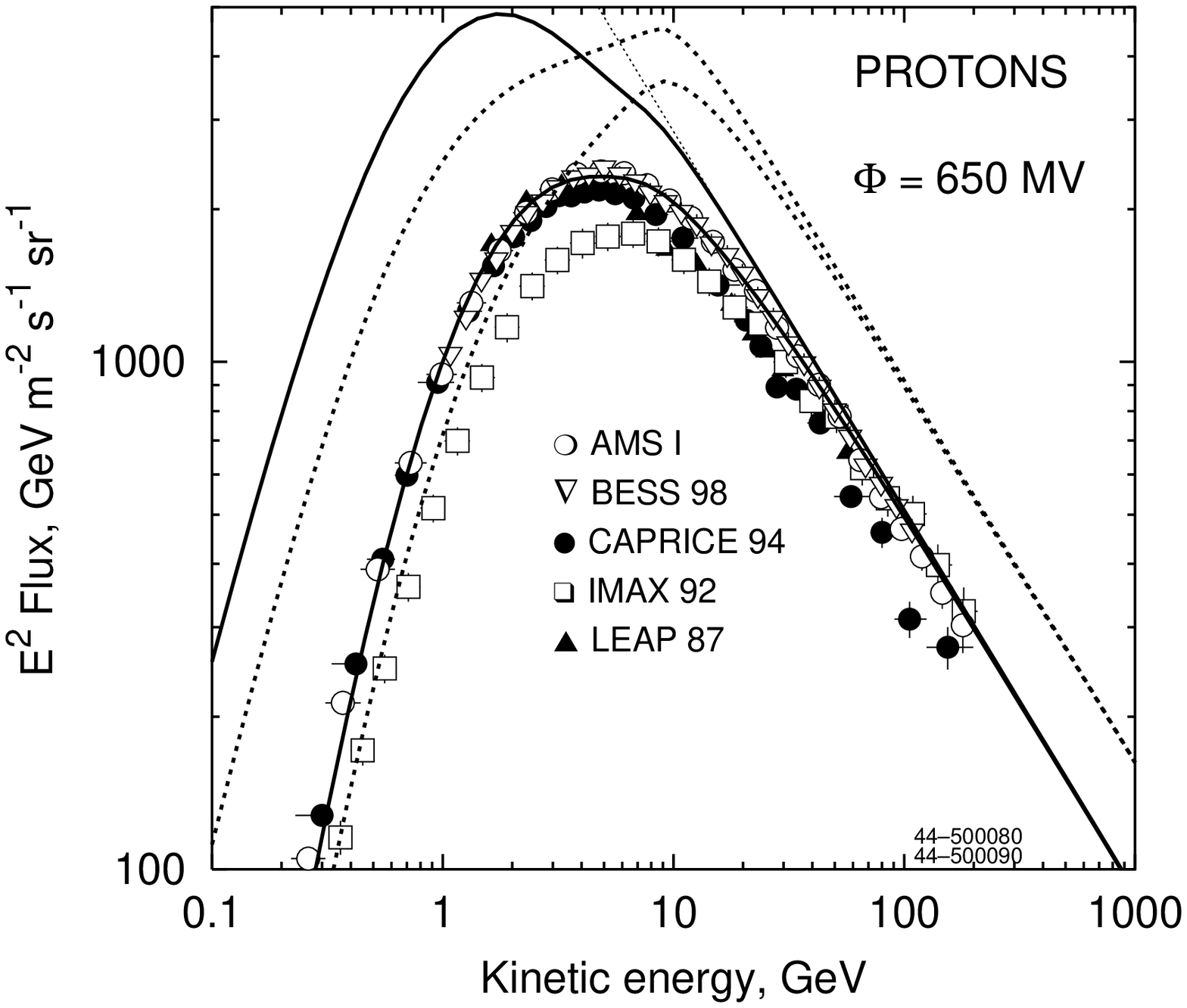,height=15pc}
            \psfig{figure=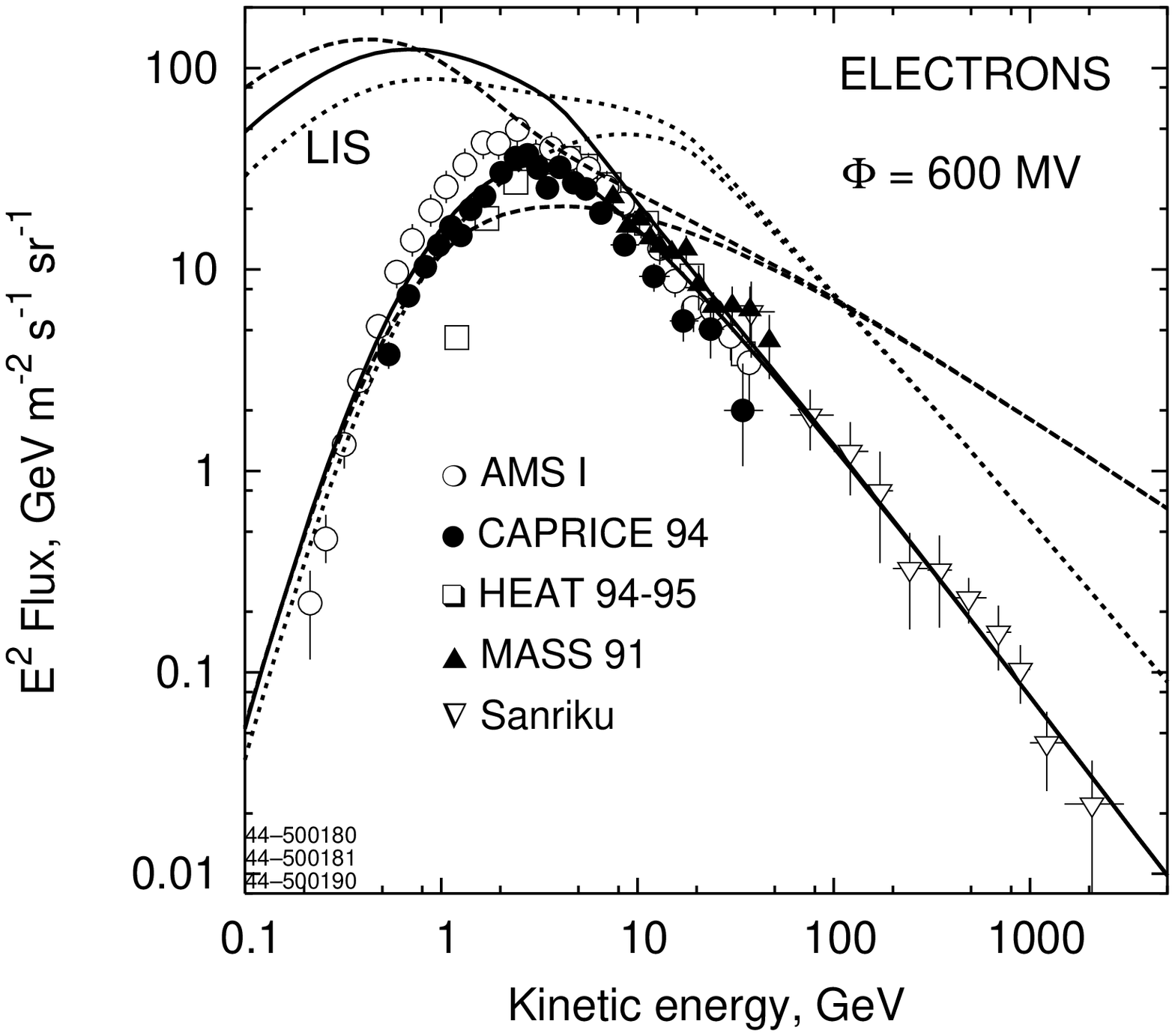,height=15pc}}
\caption{ Directly-observed (solid), and modified (dots)  proton and electron spectra.
The modified spectra are deduced from fits to antiproton and  \gray data 
\cite{2004ApJ...613..962S}. 
LIS marks local interstellar spectra.
Data shown are from AMS01, BESS, CAPRICE94, IMAX, LEAP, HEAT, MASS91, SANRIKU.
}
\label{protons_electrons_SMR2004}
\end{figure}

\begin{figure}
\centerline{\psfig{figure=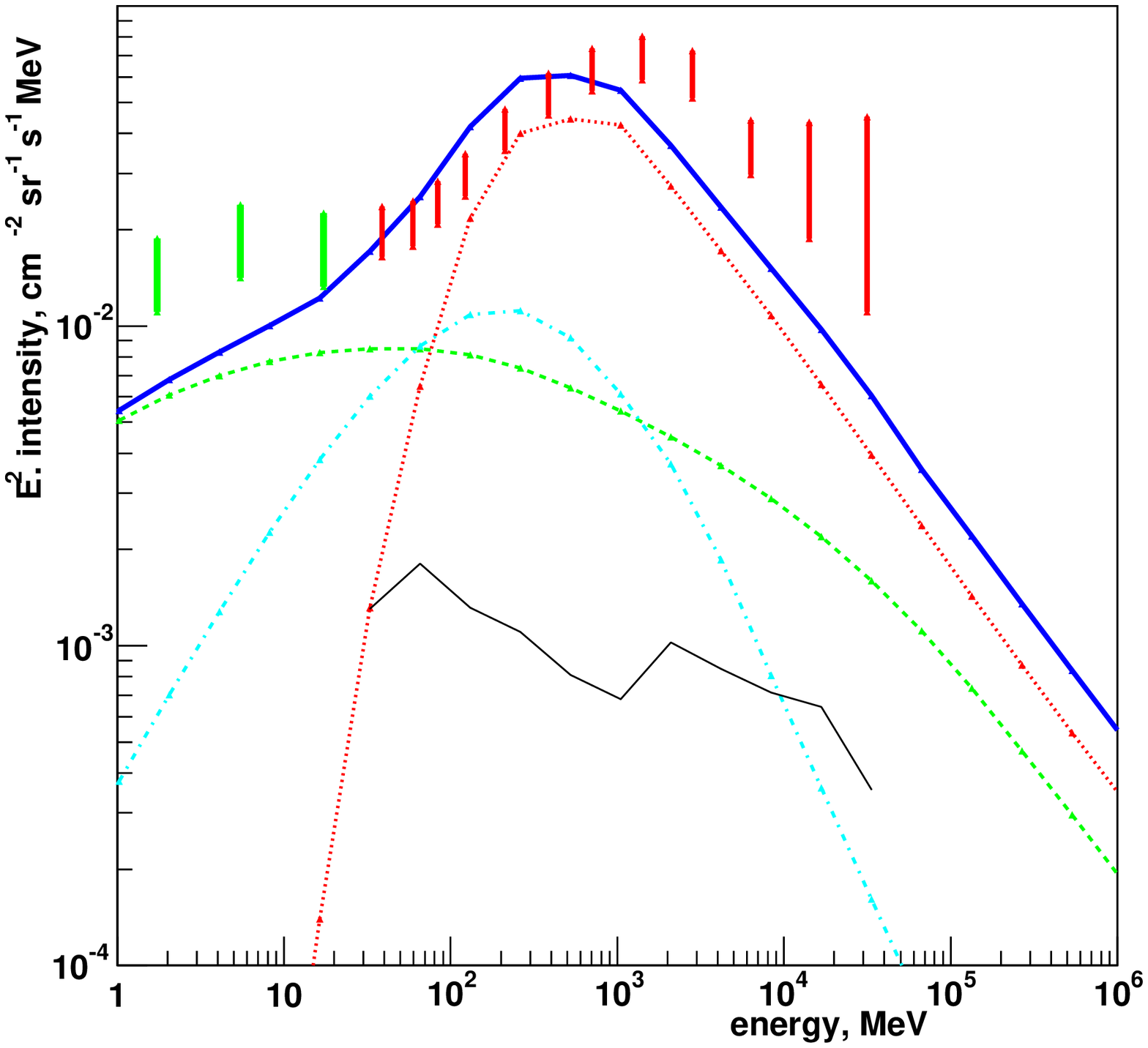,height=15pc}
            \psfig{figure=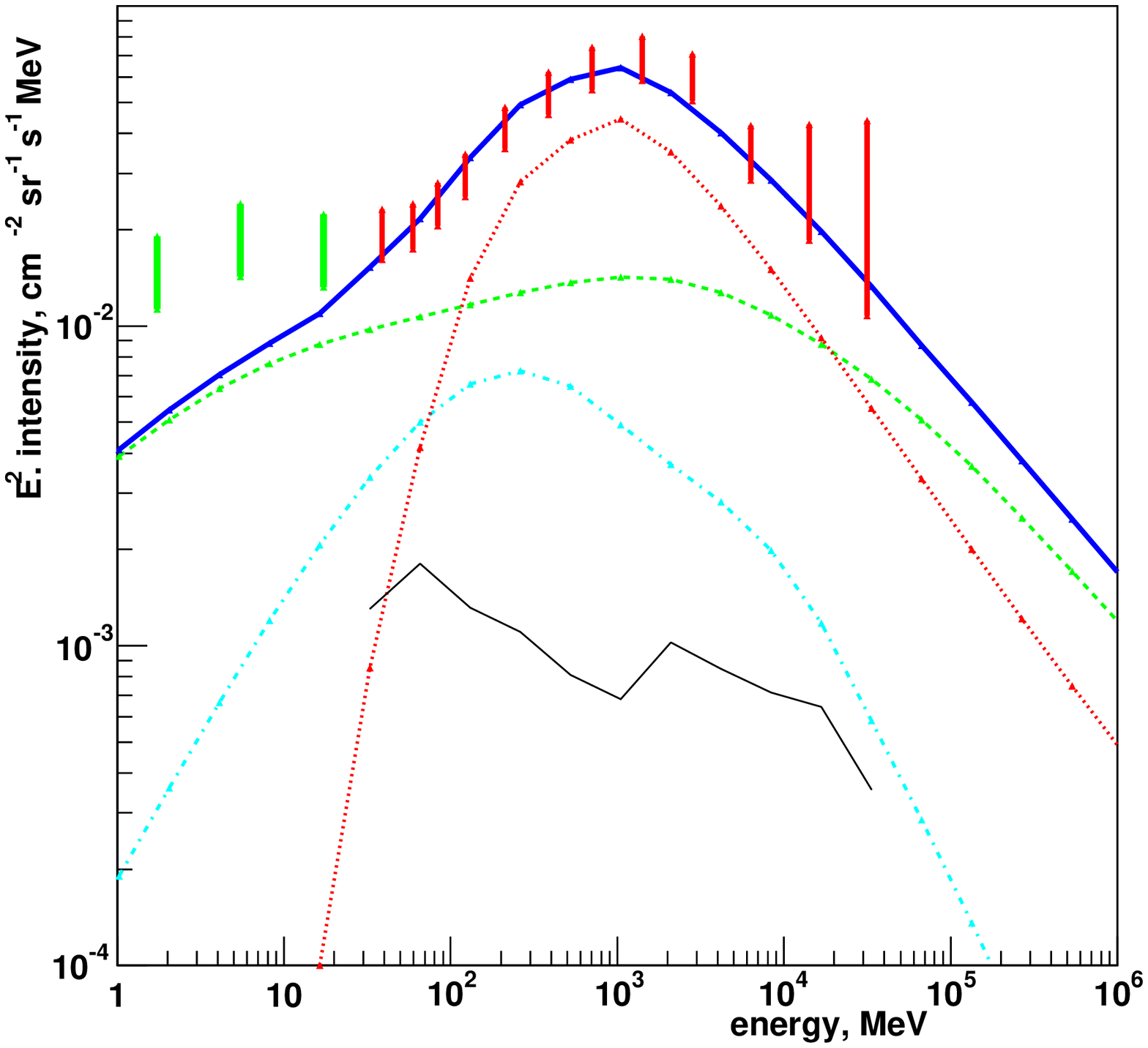,height=15pc}}
\caption{\gray spectrum of inner Galaxy ($330^\circ<l<30^\circ, |b|<5^\circ$) for model based on the directly-observed CR spectra and modified spectra shown in
Fig.~\ref{protons_electrons_SMR2004}.
Red bars:  EGRET data, including points above 10 GeV, see
 \cite{2005A&A...444..495S}. 
Green bars: COMPTEL.
Light blue line: bremsstrahlung, green line: inverse Compton scattering, red line: $\pi^0$-decay, black line: extragalactic background, dark blue line: total.
This is an update of the spectra shown in \cite{2004ApJ...613..962S}. 
}
\label{gamma_spectra_SMR2004}
\end{figure}

To illustrate the current state of the art, we show spectra and profiles from a recent GALPROP model compared with CGRO/EGRET and COMPTEL data.
This is based on
\citep{2004ApJ...613..962S} 
and
\citep{2004A&A...422L..47S} 
where full details can be found.
The first model is based simply on the directly-measured CR spectra together with a radial gradient in the CR sources.
It is immediately clear that the spectrum is not well predicted, being below the EGRET data for  energies above 1 GeV.
However  remembering that this is an {\it unfitted} prediction, it does show that the basic assumption that the \grays
are produced in CR interactions is correct; the factor 2 differences are telling us something about the remaining uncertainties.
The extent to which the CR spectra have to be modified to get a good fit including the GeV excess is shown in Fig.~\ref{protons_electrons_SMR2004}
 and the resulting \gray model prediction
in Fig.~\ref{gamma_spectra_SMR2004}. The difference between the directly-observed and modified spectra are within plausible limits considering
solar modulation and spatial fluctuations in the Galaxy on scales $>$100 pc. A detailed justification is lacking however,
so this so-called `optimized' model is just an existence proof rather than a conclusive result.
Other more drastic modifications to the CR spectrum have been proposed 
 \cite{2004cgrs.conf..279M}  
as follows:

(i) a very hard electron injection spectrum, which could be allowed invoking large fluctuations due to energy losses and the
stochastic nature of supernova remnants in space and time -- the solar vicinity is not necessarily a `typical' place for electron measurements 
to be representative
\cite{1998ApJ...507..327P,2000ApJ...537..763S}. 
 However the variations required are even larger than can  reasonably be expected
\citep{2004ApJ...613..962S}, 
so this model seems unlikely.

(ii) a hard proton spectrum, again invoking spatial variations in the Galaxy so the solar vicinity is not typical; this is more difficult
than for electrons since the proton energy-losses are negligible. It turns out this possibility can be ruled out on the basis of
antiproton measurements: too many antiprotons would be produced by the same protons which generate the gamma rays 
\citep{1998A&A...338L..75M,2004ApJ...613..962S,2004cgrs.conf..279M}. 
A related suggestion invokes the dispersion in the radio spectral indices of SNR,
which indicates a dispersion in electron indices and if
 assumed to apply to CR protons
\cite{2001A&A...377.1056B} 
could produce the GeV excess; this should also be tested against antiprotons.

Perhaps a completely different source of the excess GeV gamma rays is present, and the possibility of a dark-matter
origin has been pursued 
\cite{2005A&A...444...51D} 
but found to produces an excess of CR antiprotons
\cite{2006JCAP...05..006B}.  
 We will not enter this debate here since it is not directly related to
the problem of CR propagation. It does however show how essential is  a good understanding of the CR-induced \grays in the Galaxy
for the study of potentially  more fundamental physics
\cite{2006astro.ph..9768M}. 

\begin{figure}
\centerline{\psfig{figure=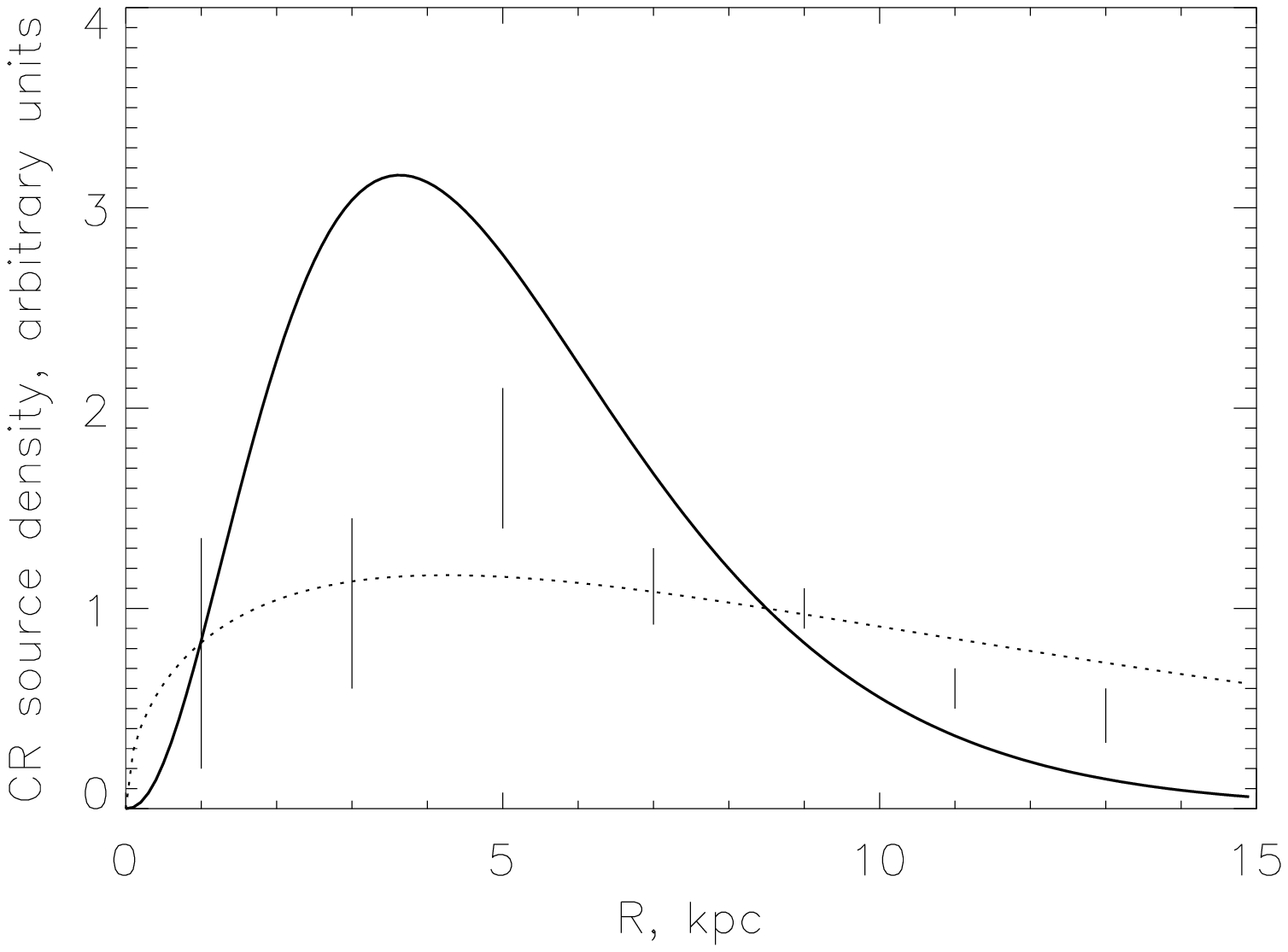,height=15pc}}
\caption{Possible CR source  distributions as function of Galactocentric radius $R$: 
 pulsars (solid line), SNR (vertical bars), from  \grays assuming constant H$_2$-to-CO relation (dotted line)
\cite{2004A&A...422L..47S}. 
}
\label{CR_source_distributions_AA2004}
\end{figure}

\begin{figure}
\centerline
{\psfig{figure=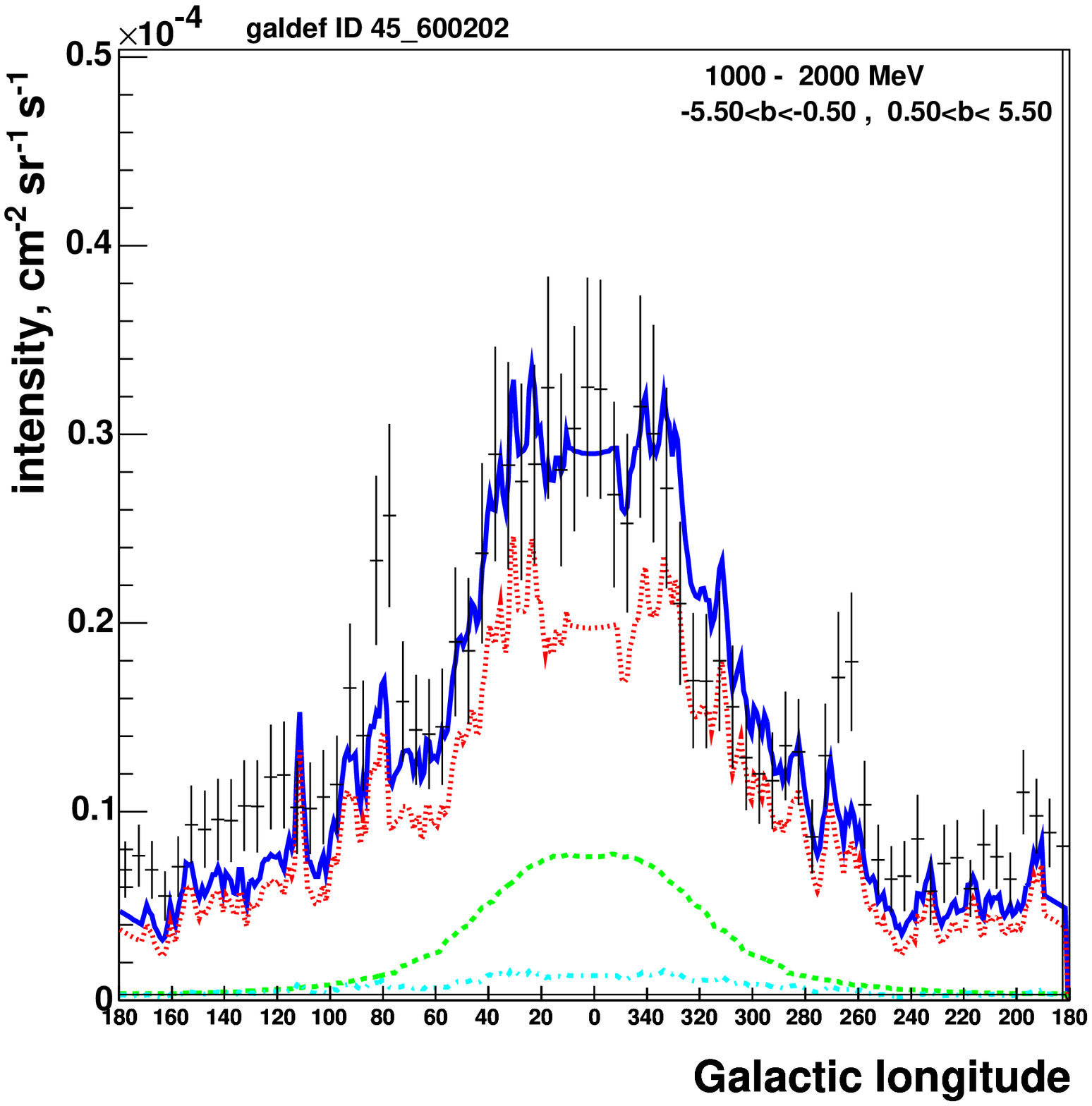,height=15pc} 
 \psfig{figure=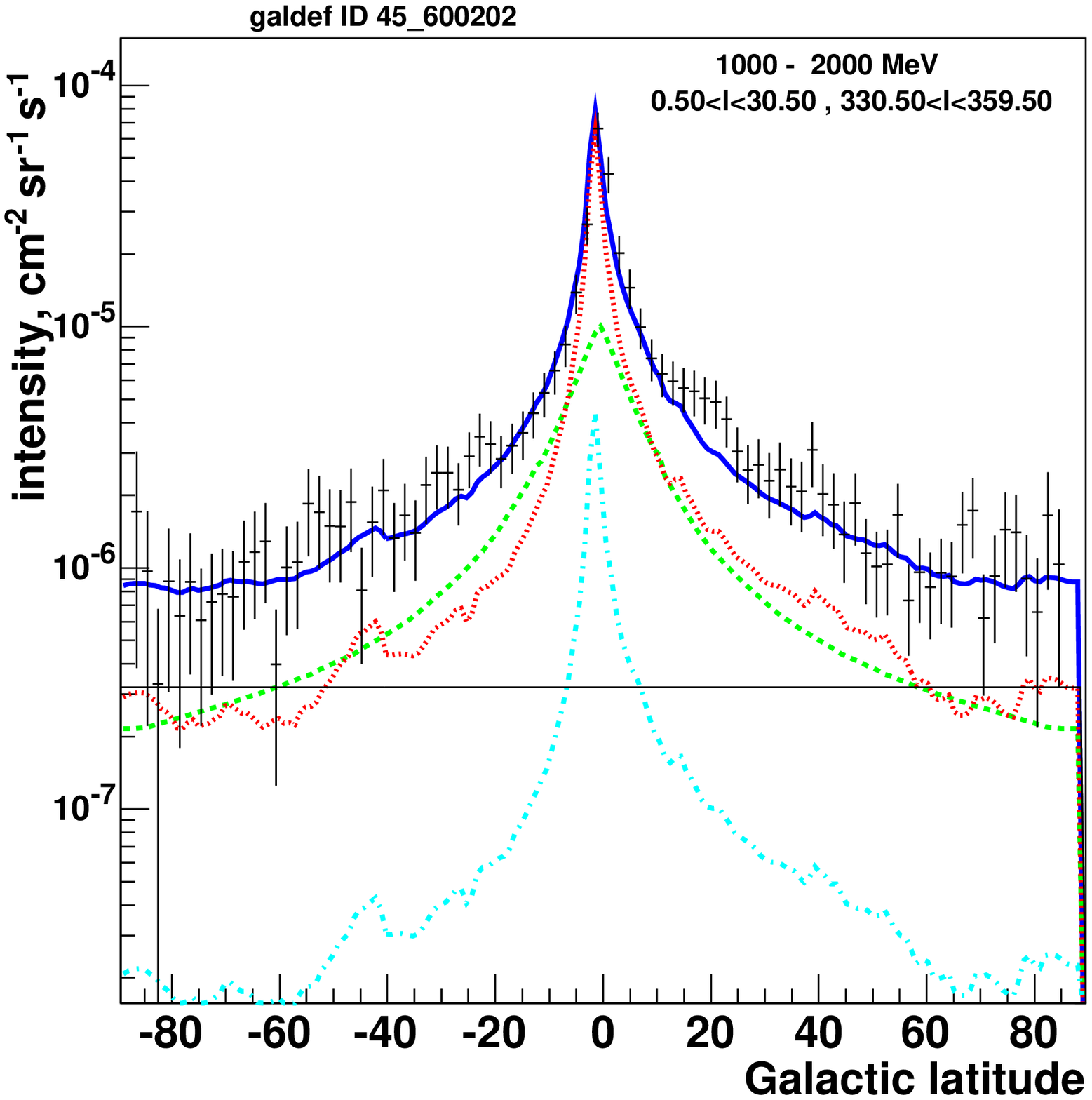,height=15pc}}
\caption{\gray longitude and latitude profiles, for model with pulsar source distribution and Xco varying with radius
\cite{2004A&A...422L..47S}. 
Light blue line: bremsstrahlung, green line: inverse Compton scattering, red line: $\pi^0$-decay, black line: extragalactic background, dark blue line: total.
}
\label{gamma_profiles_AA2004}
\end{figure}

 The angular distribution of \grays provides an essential test for any model. The problem
with the large expected gradient from SNR is critical.  The distribution
of SNR is hard to measure because of selection effects, so this problem could be safely ignored in the past,
but now the distribution of pulsars can be determined with reasonable accuracy
and this should trace SNR since supernovae are pulsar progenitors. The pulsar gradient is indeed larger than originally deduced from  \grays
as shown in Fig.~\ref{CR_source_distributions_AA2004}.
The distribution of SNR in external galaxies shows a similar gradient to pulsars
 \citep{2004Ap&SS.289..283S}. 
The problem therefore
remains, but a possible solution may be found in another  uncertainty, the
Galactic distribution of molecular hydrogen. Since we have to rely on the
CO tracer molecule, any variation in the relation of CO to H$_2$ will affect
the interpretation of the \gray data. 
\citep{2004A&A...422L..47S} 
noted that there is independent evidence for an increase in the ratio of  H$_2$ to CO
with Galactocentric radius, related to Galactic metallicity gradients,
and that this can resolve the problem, allowing CR sources to follow the SNR distribution as traced by pulsars.
Fig.~\ref{gamma_profiles_AA2004} shows longitude and latitude profiles based on such a model, showing a satisfactory agreement with EGRET data.
However the magnitude of the variation in H$_2$ to CO, or its relation to metallicity and other
effects, is uncertain so the issue will need more study. Future data from GLAST will help by
giving much finer angular resolution and the possibility of better separating the molecular and atomic
hydrogen components.
 Another possible solution to the gradient puzzle in terms of a radially dependent Galactic wind was proposed by
 \citep{2002A&A...385..216B} 
(see Sections 2.4 and 2.12).

Note the large contribution  inverse-Compton emission to the spectra and profiles (Figs.~\ref{gamma_spectra_SMR2004}, \ref{gamma_profiles_AA2004});
this is the reason that the predicted emission is in good agreement with the EGRET data
right up to the highest latitudes (where the gas-relation pion-decay emission is small).
Therefore \grays constrain both protons and electrons, the different angular distributions being clearly distinguishable.
The high-latitude inverse-Compton emission is independent evidence for the existence of a CR halo extending
up to several kpc above the plane, deduced from radioactive CR isotopes as explained in Section 3.2.
Interestingly, {\it secondary electrons and positrons} make a significant contribution to the lower-energy 
bremsstrahlung and inverse-Compton gamma-rays
\cite{2004ApJ...613..962S};  
so we can in principle observe secondaries
from all over the Galaxy, complementary to the local direct measurements.



\subsection{Antiprotons and positrons}

The spectrum and origin of antiprotons in CR has been a matter of active
debate since the first reported detections in balloon flights
\cite{1979PhRvL..43.1196G,1985ICRC....2..362B}.  
There is a consensus that most of
the CR antiprotons observed near the Earth are secondaries
\cite{1996PhRvL..76.3057M}. 
Due to the kinematics of secondary production, the
spectrum of antiprotons has a unique shape distinguishing it from
other cosmic-ray species. It peaks at about 2 GeV, decreasing sharply
towards lower energies. In addition to secondary antiprotons there
are possible sources of primary antiprotons; those most often
discussed are  dark matter particle annihilation and evaporation of
primordial black holes.

In recent years, new data with large statistics on both low and high
energy antiproton fluxes have become available
\cite{1996ApJ...467L..33H,2002AA...392..287G,2000PhRvL..84.1078O,2000ApJ...534L.177B,2000ApJ...545.1135S,2001APh....16..121M,2002PhRvL..88e1101A} 
thanks mostly to continuous improvements of the BESS instrument and its
systematic launches every 1-2 years. This allows one
to test models of CR propagation and heliospheric modulation.
Additionally, accurate
calculation of the secondary antiproton flux predicts the  ``background''
for searches for exotic signals such as WIMP annihilation. 

Despite numerous efforts and overall agreement on the secondary nature
of the majority of CR antiprotons, published estimates of the expected
flux differ significantly  (see, e.g., Fig.~3 in 
\cite{2000PhRvL..84.1078O}).  
Calculation of the secondary antiproton flux is a complicated task.
The major sources of uncertainties are three-fold: (i) incomplete
knowledge of cross sections for antiproton production, annihilation,
and scattering, (ii) parameters and models of particle propagation in
the Galaxy, and (iii) modulation in the heliosphere. While the
interstellar antiproton flux is affected only by uncertainties in the
cross sections and propagation models, the final comparison with
experiment can only be made after correcting for the solar
modulation. Besides, the spectra of CR nucleons have been directly
measured only inside the heliosphere while we need to know the
spectrum outside, in interstellar space, to compute the antiproton
production rate correctly.

It has  recently been shown
\cite{2002ApJ...565..280M} 
 that accurate antiproton
measurements during the last solar minimum 1995--1997
\cite{2000PhRvL..84.1078O} 
 are inconsistent with existing propagation models
at the $\sim$40\% level at about 2 GeV, while the stated measurement
uncertainties in this energy range are now $\sim$20\%.
The conventional models based on local CR measurements,
simple energy dependence of the diffusion coefficient, and uniform CR
source spectra throughout the Galaxy fail to reproduce
simultaneously both the secondary to primary nuclei ratio
and antiproton flux.

A reacceleration model designed to match secondary to primary nuclei
ratios  produces too few antiprotons because the diffusion coefficient is too large.
 Models without reacceleration can
reproduce the antiproton flux, but cannot explain the
low-energy decrease in the secondary-to-primary nuclei ratios.  To be
consistent with both, the introduction of breaks in the
diffusion coefficient and the injection spectrum is required, which
would suggest new phenomena in particle acceleration and propagation.
A solution in terms of propagation models requires a break in the
diffusion coefficient at a few GeV
which  has been interpreted as a
change in the propagation mode
\cite{2002ApJ...565..280M}. 
The latter  calculation  employs a modern steady-state drift model of
propagation in the heliosphere to predict the proton and
antiproton fluxes near the Earth for different modulation levels and
magnetic polarity.

If our local environment influences the spectrum of CR, then
it is possible to solve the problem by
invoking a fresh ``unprocessed'' nuclei component
at low energies 
\cite{2003ApJ...586.1050M},  
 which may be produced in the Local Bubble.
The idea is that primary CR like C and O have a local low-energy component,
while secondary CR like B are produced Galaxy-wide over
the confinement time of 10--100 Myr.
In this way an excess of B, which appears when propagation
parameters are tuned to match the $\bar p$ data, can be
eliminated by an additional local source of C (see Section 3.1). 
The model appears to be able to describe a variety of
CR data, but at the cost of additional parameters.

A consistent $\bar p$ flux in reacceleration
models can be obtained if there are additional sources of  protons
$\hbox{\rlap{\hbox{\lower3pt\hbox{$\sim$}}}\lower-2pt\hbox{$<$}}$20 GeV
\cite{2003ICRC....4.1921M}. 
This energy is above the
$\bar p$ production threshold and effectively produces $\bar p$'s
at $\hbox{\rlap{\hbox{\lower3pt\hbox{$\sim$}}}\lower-2pt\hbox{$<$}}$2 GeV.
The intensity and spectral shape of this
component could be derived by combining restrictions from $\bar p$'s
and diffuse $\gamma$-rays.

It is clear that accurate measurements of $\bar p$ flux
are the key to testing current propagation models.
If new measurements confirm the $\bar p$ ``excess,'' current
propagation and/or modulation models will face a challenge.
If not -- it will be evidence that the reacceleration model is
currently the best one to describe the data.
Here we have to await for the next BESS-Polar flight which should bring
much larger statistics, and Pamela 
 \cite{2006astro.ph..8697P},
currently in orbit.

Positrons in CR were discovered in 1964 
\cite{1964PhRvL..12....3D}. 
The ratio of positrons to  electrons is approximately 0.1
above $\sim$1 GeV as measured near the Earth. Secondary positrons
in CR are mostly the decay products of charged pions and kaons ($\pi^+$, K$^+$)
produced in  CR  interactions with  interstellar gas.
The calculation of the positron production by CR generally
agrees with the data and indicates that the majority of CR
positrons are secondary
\cite{1982ApJ...254..391P,1998ApJ...493..694M}.  
Some small fraction of positrons may also come from
 sources
 \cite{1999APh....11..429C}, 
such as pulsar winds
\cite{1996ApJ...459L..83C} 
 or WIMP annihilations
\cite{1999PhRvD..59b3511B}. 

In interstellar space the secondary positron flux below $\sim$1 GeV
is comparable
 to the electron flux which makes positrons non-negligible
contributors to the diffuse $\gamma$-ray emission in the MeV range
 \cite{2004ApJ...613..962S}.  
The contribution of secondary positrons and electrons
in  diffuse Galactic  $\gamma$-ray emission models essentially improves
the agreement with the data,  making secondary positrons effectively
detectable in $\gamma$-rays.

The accuracy of  spacecraft and balloon-borne experiments
has substantially increased during the last decade. The data taken
by various instruments including AMS, CAPRICE and HEAT 
\cite{2002PhR...366..331A,2006astro.ph..5254G,2000ApJ...532..653B,2001ApJ...559..296D}       
 are in agreement
with each other within the error bars, which are however still quite large.
A recent measurement of CR positron fraction
by the HEAT instrument
\cite{2004PhRvL..93x1102B} 
 indicates an excess (feature)
between 5 and 7 GeV at the level of $\sim$3$\sigma$ over the
smooth prediction of a propagation model 
\cite{1998ApJ...493..694M}. 
When all the HEAT flights are combined, the data show an
excess above $\sim$8 GeV, with the most significant point at $\sim$8 GeV
\cite{2004PhRvL..93x1102B}. 
The Pamela satellite 
\cite{2006astro.ph..8697P} 
currently in orbit, will have a very good positron statistics.


\subsection{Electrons and  Synchrotron Radiation}

CR electrons require a separate treatment from nuclei because of their rapid energy losses
and their link with synchrotron radiation.
Their low energy density compared with nuclei ($\approx 1\%$) is not yet understood;
standard SNR shock acceleration does not predict such a ratio and it is normally
a free parameter in models e.g.
\cite{2001SSRv...99..305E}. 
Direct measurements extend from MeV to TeV ;
at low energies solar modulation is so large that the interstellar fluxes
are really unknown. At TeV energies the statistics are very poor
but new experiments are in progress.
Young SNR contain  TeV electrons as is shown by their non-thermal X-ray emission,
and energies up to 100 TeV are possible
\cite{1999ICRC....3..480A}. 
The rapid energy losses mean that contributions from old nearby SNR such as Loop 1, Vela, the Cygnus Loop and MonoGem
could make an important contribution to the local electron spectrum above 100 GeV and  dominate above 1 TeV,
see Fig.~\ref{electrons_Kobayashi}
\cite{2004ApJ...601..340K}. 
The rapid energy losses of electrons also produce time- and space-dependent effects described in the next section.
The essential propagation effects are seen in Fig.~\ref{protons_electrons_SMR2004};
at energies below 1 GeV the interstellar spectrum is flatter than the injection spectrum, due to Coulomb losses;
there is an intermediate energy range where the spectrum is similar to that at injection,
and at high energies it steepens due to inverse Compton and synchrotron losses and
energy-dependent diffusion.
For propagation models we need an injection spectrum which  reproduces the observations;
a typical spectral index is 2.4, similar to  nuclei, which produces the observed high-energy slope
of 3.3 
\citep{2004ApJ...613..962S}. 
The synchrotron spectral index is $\beta_\nu=(\gamma - 1)/ 2$ for an electron index $\gamma$;
the observed   $\beta_\nu=0.6 - 1$, increasing with frequency,  implies $\gamma=2.4 - 3$ increasing with energy, in satisfactory agreement.
Detailed geometrical models of the Galactic synchrotron emission have been constructed
 \cite{1985A&A...153...17B,1990ICRC....3..229B}. 
A more physical interpretation requires propagation and magnetic field modelling as for example in
\cite{2000ApJ...537..763S}. 
Combined with \gray data this allows the magnetic field to be determined as a function of
Galactocentric radius independent of other techniques
\cite{2000ApJ...537..763S}. 
Since that work a great deal of new radio data has become available both  from new ground-based surveys
and from balloons and satellites for cosmology (e.g. WMAP). Since Galactic synchrotron is an
essential `foreground' for CMB studies there is a large overlap of interest between these fields.
Exploitation of these data for CR studies  has only just  begun but is expected to accelerate
with the forthcoming Planck mission.

Radio continuum observations of other galaxies provide a complementary view of electrons via synchrotron radiation
\cite{2005mpge.conf..193B,2006astro.ph..3531B}. 
The classical case is the edge-on galaxy NGC891
\cite{1978A&A....62..397A,2006ApJ...647.1018H} 
 which has a non-thermal halo extending to several kpc,
and this helped to give credence to the idea of a large halo in our Galaxy.
More recent observations of this and other edge-on galaxies confirm large halos
\cite{1998LNP...506..555D}. 
The spectral index variations due to energy losses provide a test of propagation in principle, although this
has not been very fruitful up to now. This effect can also be measured  in starburst galaxies 
\cite{2005mpge.conf..156H}. 
Face-on galaxies like  M51 show marked spiral structure in the continuum
\cite{2005mpge.conf..193B}, 
but this is probably mainly a magnetic-field effect.
It is not possible reliably to separate CR and magnetic field variations, although this is often attempted
assuming equipartition of energy between CR and field,
 which although having no firmly established physical basis nevertheless gives field values quite similar to other methods
\cite{2005AN....326..414B,2006astro.ph..3531B}. 

The tight radio-continuum/far-infrared correlation for galaxies
\cite{1997A&A...320...54N,2005A&A...437..389M,2006A&A...456..847P} 
 is worth mentioning here.
It holds both within galaxies
(down to 100 pc or less) 
 and from galaxy to galaxy, and over several orders of magnitude in luminosity;
it presumably contains  clues to CR origin and propagation.
The `CR calorimeter' 
\cite{1989A&A...218...67V} 
is the simplest interpretation.
Simply relating both CR production  and UV (reprocessed to far-infrared) to star formation rates is apparently insufficient
to explain the close correlation, and this  has given rise to several different  interpretations including  hydrostatic regulation,  without being very conclusive as yet
\cite{1997A&A...320...54N,2005A&A...437..389M,2006A&A...456..847P}. 


\subsection{Time- and space-dependent effects}

CR propagation is usually modelled as a spatially smooth, temporally steady-state process.
Because of the rapid diffusion and long containment time-scales in the Galaxy this is usually a sufficient approximation,
but there are cases where it breaks down.
The rapid energy loss of electrons above 100 GeV  and the stochastic nature of their sources produces spatial and temporal
variations.  Supernovae are stochastic events
and each SNR source of CR lasts only perhaps $10^4 - 10^5$ years, so that they leave their
imprint on the distribution of electrons. This leads to
large fluctuations in the CR electron density at high energies, so that
the electron spectrum measured near the Sun may not be typical
\cite{2004ApJ...613..962S}; 
a statistical calculation of this effect can be found in 
\cite{1998ApJ...507..327P,2001ICRC....5.1964S} 
and for local SNR in
\cite{2004ApJ...601..340K}. 
These effects are  much smaller for nucleons since there are essentially no energy losses,
but they can still be important
\cite{2002JPhG...28..359E};  
for the theory of CR fluctuations for a galaxy with random SNR events see
\cite{2006AdSpR..37.1909P} 
and Section 3.5.
Such effects can influence the B/C ratio 
\cite{2004ApJ...609..173T,2005ApJ...619..314B}  
mainly through variations in the primary spectra.
The local bubble can also have an effect on the energy-dependence of B/C
\cite{2003ApJ...586.1050M}. 
Dispersion in CR injection spectra from SNR may cause the locally observed spectrum to deviate from the average
\cite{2001A&A...377.1056B}. 


\section{Summary Points list}

Progress in CR research is rapid at the present time with the new generation of detectors
for both direct and indirect measurements. But still it is hard to pin-down particular theories
and even now the origin of the nucleonic component is not proven 
\cite{2006A&A...449..223A}, 
although SNR are the leading candidates.

The main points we want to make are
\begin{itemize}
\item the importance of considering all relevant  data, both direct (particles) and indirect ($\gamma$-ray, synchrotron) measurements
\item increase in computing power has made many of the old approximations for interpreting CR data unnecessary
\item new high-quality data will require detailed numerical  models
\end{itemize}

\section{Future Issues} 
We end by listing some of the open questions regarding CR propagation  which might be answered with  observations in the near future.
\begin{itemize}

\item the interpretation of the energy-dependence of the secondary/primary ratio, requiring accurate measurements at both, low and high energies
\item the size of the propagation region -- existence of an extended halo, requiring measurements of radioactive species over a broad energy range
\item the relative roles of diffusion, convection and reacceleration 
\item the importance of local sources to the primary CR flux
\item the origin of the GeV excess in \grays relative to the prediction based on locally observed CR
\item the CR source distribution: is it like SNR or not? 
\item are positrons and antiprotons explained as secondaries from primary CR or is there a -- perhap exotic -- excess? 
\end{itemize}

In addition, on the theoretical side, we mention
\begin{itemize}

\item the relation of CR-dynamical  models of the Galaxy to  CR propagation theory

\end{itemize}


\section{Related Resources}
GALPROP website: {\tt http://galprop.stanford.edu},\\
GLAST website: {\tt http://glast.stanford.edu}


\section{Key terms, definitions and acronyms}
Key terms/definitions:

Direct measurements: measurements on CR themselves with detectors in balloons and satellites.
Indirect measurements: via electromagnetic radiation (\grays and synchrotron) emitted by CR interacting with interstellar matter and radiation fields.
Primary CR: CR accelerated at sources (for example in SNR).
Secondary CR:  CR generated by spallation of primary CR on interstellar gas.
$p$: particle momentum,
$R$: particle magnetic rigidity = $pc/Ze$,
$\beta$: velocity/speed-of-light,
$z_h$: height of CR halo in direction perpendicular to Galactic plane

Acronyms:

ACE: Advanced Composition Explorer,
B/C:  cosmic-ray Boron-to-Carbon ratio,
CMB: cosmic microwave background,
CGRO: Compton Gamma Ray Observatory,
CR: cosmic rays,
EGRET: Energetic gamma-ray telescope,
GALPROP: Galactic Propagation code,
GLAST: Gamma-ray Large Area Telescope,
H.E.S.S.: High Energy Stereoscopic System,
ISRF: interstellar radiation field,
PLD: path-length distribution,
SNR: supernova remnant

\section{Acknowledgements}
I.V.M. acknowledges partial support from a NASA Astronomy and Physics Research and Analysis Program (APRA) grant.


\bibliographystyle{arnuke}  

\bibliography{1990AdSpR__10___69H,strong,2000A+A___354__423L,2000A+A___358L__79V,extra,ISSI,SNR,knee,Diffusion,anisotropy}



\end{document}